\documentclass[preprint,authoryear,11pt]{elsarticle}

\makeatletter
\def\ps@pprintTitle{%
  \let\@oddhead\@empty
  \let\@evenhead\@empty
  \let\@oddfoot\@empty
  \let\@evenfoot\@oddfoot
}
\makeatother

\usepackage{amsfonts,amsmath,amstext,amssymb,amsthm}
\usepackage{mathtools} 

\usepackage[usenames,dvipsnames]{xcolor} 

\usepackage{caption}
\usepackage{subcaption}

\newcommand{\prob}[1]{\mathbb{P}\left(#1\right)}

\def\E{\mathbb{E}\,}

\newcommand{\bankEquity}[1]{e_{#1}}
\newcommand{\investorEquity}[1]{w_{#1}}
\newcommand{\bankBonds}[1]{x_{#1}}
\newcommand{\investorBonds}[1]{z_{#1}}
\newcommand{\deposits}[1]{y_{#1}}
\newcommand{\price}[1]{\pi_{#1}}

\newcommand{\priceDemandLessThanMax}{\widetilde{\pi}}

\newcommand{\reserves}[1]{r_{#1}}
\newcommand{\currency}[1]{c_{#1}}
\newcommand{\currencyFraction}{\mu}

\newcommand{\bankDemand}[1]{D_\text{\it b}({#1})}
\newcommand{\investorDemand}[1]{D_\text{\it i}({#1})}
\newcommand{\totalDemand}[1]{D_\text{total}({#1})}
\newcommand{\CARconstraint}[1]{{C}_\text{cap.\ req.}({#1})}
\newcommand{\CARconstraintBracket}[1]{{C}_\text{cap.\ req.}[{#1}]}

\newcommand{\ve}[2]{v_{{#2}}}

\DeclareMathOperator*{\argmax}{arg\,max}

\newtheorem{example}{Example}
\newtheorem{definition}{Definition}
\newtheorem{lemma}{Lemma}
\newtheorem{assumption}{Assumption}

\begin{document}

\begin{frontmatter}

\title{Inside Money, Procyclical Leverage, and Banking Catastrophes}


\author[cdb1,cdb2]{Charles D.~Brummitt\corref{mycorrespondingauthor}}
\address[cdb1]{Department of Mathematics, University of California, Davis, CA 95616 USA}
\address[cdb2]{Complexity Sciences Center, University of California, Davis, CA 95616 USA}
\cortext[mycorrespondingauthor]{Corresponding author}
\ead{cbrummitt@math.ucdavis.edu}

\author[rs1,rs2]{Rajiv Sethi}
\address[rs1]{Department of Economics, Barnard College, Columbia University, New York, NY 10027}
\address[rs2]{Santa Fe Institute, Santa Fe, NM 87501 USA}
\ead{rs328@columbia.edu}

\author[djw1]{Duncan J.\ Watts}
\address[djw1]{Microsoft Research, New York, NY 10011 USA}
\ead{duncan@microsoft.com}

\begin{abstract} 

We explore a model of the interaction between banks and outside investors in which the ability of banks to issue inside money (short-term liabilities believed to be convertible into currency at par) can generate a collapse in asset prices and widespread bank insolvency. The banks and investors share a common belief about the future value of certain long-term assets, but they have different objective functions; changes to this common belief result in portfolio adjustments and trade. Positive belief shocks induce banks to buy risky assets from investors, and the banks finance those purchases by issuing new short-term liabilities. Negative belief shocks induce banks to sell assets in order to reduce their chance of insolvency to a tolerably low level, and they supply more assets at lower prices, which can result in multiple market-clearing prices. A sufficiently severe negative shock causes the set of equilibrium prices to contract (in a manner given by a cusp catastrophe), causing prices to plummet discontinuously and banks to become insolvent. Successive positive and negative shocks of equal magnitude do not cancel; rather, a banking catastrophe can occur even if beliefs simply return to their initial state. Capital requirements can prevent crises by curtailing the expansion of balance sheets when beliefs become more optimistic, but they can also  force larger price declines.  Emergency asset price supports can be understood as attempts by a central bank to coordinate expectations on an equilibrium with solvency.

\medskip 


\end{abstract}

\end{frontmatter}


\section{Introduction}

Despite a large and rapidly growing theoretical literature, the mechanisms that give rise to periodic asset price booms and crises remain imperfectly understood. In this paper, we investigate one such mechanism that arises from the interaction between banks and outside investors, in which the capacity of banks to issue inside money---short-term liabilities that are believed to be convertible into currency at par---plays a central role. 
The model generates a positive correlation between asset prices and bank leverage, which is consistent with the empirical observation that banks, along with other financial intermediaries, tend to finance more of their risky, long-term assets with short-term debt when asset prices rise and less when asset prices fall~\citep{Adrian2010}. That is, bank leverage is procyclical, in contrast with leverage among households and non-financial firms, which is countercyclical. Because a static balance sheet has countercyclical leverage (i.e., a rise in asset prices mechanically lowers leverage, and vice versa), the procyclical leverage of financial intermediaries suggests that they aggressively expand their balance sheets and debt levels when asset prices are rising, and that they contract balance sheets when asset prices are declining. 

This rapid expansion and contraction of banks' balance sheets is feasible because the short-term liabilities of banks are viewed by non-banks as close substitutes for currency. The banking sector as a whole can increase its net holdings of risky assets because non-banks are willing to accept these liabilities in exchange for riskier assets~\citep{Gorton2012_book}. This observation---that the sizes of bank balance sheets can be considerably elastic---is a key element in the model explored here. Another important element is that the objective functions of banks and outside investors differ: we assume that banks operate in a manner that is risk neutral subject to an insolvency constraint, while investors are risk averse. Both groups share a common belief about the future value of certain long-term, risky assets, and they interpret new information about the future value of these assets in the same way. However, because their objectives differ, changes to the common belief about the assets' future value results in trade between the two groups.

For simplicity, we model banks and investors at the sectoral level by considering one consolidated balance sheet for each group. Initially, these two balance sheets are mutually consistent in the sense that neither group wishes to adjust its portfolio at the prevailing prices (given the commonly held belief about the future value of the assets). Whenever this belief changes, trading restores consistency of the balance sheets. In the case of a positive shock to the  commonly held belief (representing greater optimism about the future value of the assets), both groups demand more assets at the current prices. Given that the asset supply is fixed in the short term, a positive shock causes prices to rise and portfolios to be adjusted. There exists a unique price at which balance sheet consistency is restored, with assets sold by investors to banks and new short-term liabilities (inside money) created. As a result, banks expand their balance sheets and (typically) increase their leverage.

The response to a negative belief shock is not symmetric. After a negative shock, both banks and investors wish to sell assets at the prevailing prices, but there can be more than one price at which balance sheet consistency is restored. This multiplicity of equilibria results from the bank demand increasing in the asset price: lower prices correspond to greater balance sheet stress and a greater desire to sell. A sufficiently strong negative shock can eliminate two of the equilibria, leaving only one equilibrium, at which the banking sector is insolvent.\footnote{In the language of catastrophe theory~\citep{Rosser2000}, the set of equilibrium prices undergoes a fold catastrophe as the negative shock becomes more severe. More generally, varying the hypothetical shock using two parameters (rather than one) leads to a cusp catastrophe, as illustrated in Fig.~\ref{fig:negativeshock}(E) below. A qualitatively similar effect arises in models of stock market crashes induced by dynamic hedging~\citep{Gennotte1990}.} In short, the unwillingness of outside investors to absorb all the assets that banks wish to sell can cause a discontinuous decline in asset prices and widespread bank insolvency.

We show that successive positive and negative shocks of equal magnitude do not cancel; rather, a banking catastrophe can occur even if beliefs simply return to their initial state. 
In short, the model is path dependent. A system that can absorb a negative shock of a given size (without suffering a crisis) may be unable to do so if the negative shock is preceded by a positive shock. Optimism induces a buildup of leverage, which increases fragility and can lead to insolvency if the optimism is subsequently reversed. 

Finally, within the context of this model we consider the effects of policies such as emergency asset price supports and capital requirements. Support of asset prices by the central bank can coordinate beliefs on an equilibrium that is consistent with bank solvency in the face of asset price declines. Such support was accomplished by the Federal Reserve using a variety of temporary facilities created in 2008; these facilities allowed for the lending of reserves or Treasury securities to financial intermediaries in exchange for much riskier and less liquid collateral. In contrast to emergency policies that deal with an ongoing crisis, capital requirements aim to prevent crises by constraining leverage growth during asset price booms. This constraint can prevent asset price collapses and widespread insolvency when optimism is reversed; however, it can also induce greater selling in a declining market, suggesting the need for graduated or countercyclical constraints on leverage, as has been proposed by the~\cite{Longworth2010},~\cite{Goodhart2010}, and~\cite{Admati2013}.

\section{Literature}

A substantial literature has examined the buildup of fragility in the banking system during periods of low volatility. Long before the most recent crisis, \cite{Minsky1975, Minsky1986}, \cite{Crockett2000}, and \cite{BorioLowe2002} argued that periods of stable growth result in changes in financial practices that make subsequent instability more likely.\footnote{For instance, \citet[p.~5]{Crockett2000} argues that in contrast to the conventional wisdom,  ``it may be more helpful to think of risk as {\it increasing} during upswings, as financial imbalances build up, and {\it materializing} in recessions." Along similar lines, \citet[p.126]{Minsky1975} writes: ``Stability---even of an expansion---is destabilizing in that more adventuresome financing of investment pays off to the leaders, and others follow."} Inside money plays a central role in the theory of credit booms and busts in \cite{Dangetal2013} and \cite{GortonOrdonez2014}, where emerging doubts about the collateral that backs privately issued short-term debt can cause it to become ``information-sensitive", making investments in determining its quality profitable. Procyclical leverage has been shown to arise from standard risk management strategies based on the concept of Value-at-Risk (VaR), which amount to risk-neutral behavior subject to a constraint on the likelihood of insolvency~\citep{Brunnermeier2009, Shin2010_RiskandLiquidity, Danielsson2011}. Leverage cycles can also emerge in models of collateral equilibrium~\citep{Geanakoplos1997,Geanakoplos2003} based on heterogeneous beliefs, as in \cite{Fostel2008} and \cite{Geanakoplos2010_Leverage}. In these models, an asset price boom is associated with a distributional shift in wealth towards highly leveraged optimists, with a corresponding reversal when prices collapse.\footnote{See also~\cite{Lorenzoni2008} and~\cite{Bianchi2011}, who provide normative analyses of credit booms and busts.}

A closely related literature has studied the mechanisms by which a financial crisis spreads. 
Distress can be transmitted through knock-on contagion, in which the failure of one institution puts its counterparties in a network at risk; see \cite{Allen2000}, \cite{Freixasetal2000}, \cite{Watts2002}, \cite{Nier2007}, \cite{Gai2010}, \cite{May2010}, \cite{Battiston2012b}, and \cite{Elliott2012}, for instance. Even without any direct counterparty relationships (e.g., without any loans between banks),
the failure of an institution can damage others through the market impact of asset sales. This effect is especially strong when institutions' exposures to assets are highly correlated~\citep{Caccioli2012, Huang2013,Caballero2009}. Contagion can also occur through an information channel: distress to an asset class or to a financial institution can lead to worry about other assets and institutions~\citep{Krishnamurthy2012, Acharya2008, Anand2012}.

Our contribution to this literature is to highlight the asymmetric reaction of balance sheets to positive and negative shocks, the possibility of discontinuous asset price declines, and the path dependence of portfolio adjustments in response to changing beliefs. We make a number of simplifying assumptions that block the mechanisms that have already been extensively explored. First, we assume that the banks and investors are homogeneous, so we model them at the sectoral level as two consolidated balance sheets. Second, we consider only a single class of risky assets, which may be interpreted as securities backed by pools of other assets such as mortgage loans, which in turn are backed by real assets such as housing.  Third, we assume that the supply of these assets is fixed, effectively excluding the role of banks in underwriting new assets, a potentially important source of systemic risk. Fourth, we assume that the banks and investors share a common belief about the probability distribution governing the bonds' future value and that they update their beliefs identically in response to new information. Finally, we assume that the short-term liabilities of banks are perceived to be risk-free at all times, so they never become information-sensitive in the sense of \cite{Gorton2012_book}. This assumption rules out the possibility of runs, as in \cite{Diamond1983}, and it allows us to focus on asymmetric reactions to shocks, discontinuities in market clearing prices, and the path dependence of portfolio adjustments.

\section{The Model} \label{sec:model}

\subsection{Balance Sheets}
Our model proceeds via a sequence of unitless time steps $t \in \{0,1, \dots, T, T+1\}$. There is a single class of risky assets, henceforth called {\it bonds}, available in fixed supply. At each step $t$, let $\bankBonds{t}$ denote the number of bonds held by the banks, where each bond has unit par value (so each bond is worth at most $1$ in period $T+1$). As described in the next section, the market-clearing bond price in period $t$, denoted by $\price{t}$, is determined by the commonly-held belief at time $t$ about the bonds' future value.\footnote{For simplicity, we assume that no interest payments are made prior to maturity (i.e., these are zero-coupon bonds), so the bond price is bounded above by its par value.} The total value of the banks' bond holdings is therefore $\price{t}\bankBonds{t}$. In addition to bonds, the banks' assets also include reserves $\reserves{t}$, which include vault cash and deposits at the central bank. 

Bank liabilities consist of short-term debt that is believed by investors to be convertible into currency at par without risk. We refer to these liabilities as {\it deposits}, broadly interpreted to include shares in money market funds, which carry an implicit if not explicit government guarantee.  These deposits $\deposits{t}$ are held by outside investors.  The banks' equity at time $t$ is the residual $\bankEquity{t} \equiv \price{t} \bankBonds{t} + \reserves{t} - \deposits{t}$. The banks' consolidated balance sheet is shown in Table \ref{tab:bankbs}.

\begin{table}[htb]
\begin{center}
\begin{tabular}{ll}
{Bank assets} & {Bank liabilities} \\ \hline
bonds $\price{t} \bankBonds{t}$ & deposits $\deposits{t}$ \\
reserves $\reserves{t}$ & equity  $\bankEquity{t}$ 
\end{tabular}
\end{center}
\caption{Consolidated bank balance sheet in period $t$.}
\label{tab:bankbs}
\end{table}%

The investors also hold a number of bonds (denoted by $\investorBonds{t}$), as well as money in the form of currency (denoted by $\currency{t}$) and bank deposits $\deposits{t}$. They carry no debt, so their equity $\investorEquity{t}$ is the sum of the assets $\price{t} \investorBonds{t} + \currency{t} + \deposits{t}$. We have in mind real money investors such as pension funds, sovereign wealth funds, insurance companies and retail mutual funds. All leverage therefore exists within the banking system, 
and we ignore leveraged, non-bank investors such as hedge funds.

For simplicity, we assume that a fixed fraction $\currencyFraction$ of money in circulation is held by investors as currency, so $\currencyFraction  \equiv \currency{t} / (\currency{t} + \deposits{t})$ is a parameter of the model, which we restrict to the interval $[0,1)$. Provided $\currencyFraction > 0$, the banks cannot expand their balance sheets without limit because, as described below, they suffer a drain of reserves as they accumulate bonds and because the total quantity of high powered money (currency plus reserves) is in fixed supply, under the control of the central bank.\footnote{Our arguments do not rely on $\currencyFraction$ being fixed, only that $\currencyFraction$ be bounded away from zero at all times, so that the expansion of bank balance sheets results in a drain on reserves.} Table~\ref{tab:nonbankbs} shows the investors' balance sheet. 
\begin{table}[hbtp]
\begin{center}
\begin{tabular}{cc}
{Investor assets} & {Investor liabilities} \\ \hline
bonds $\price{t} \investorBonds{t}$& equity $\investorEquity{t}$  \\
currency $\currency{t}$ & \\
deposits $\deposits{t}$ &   
\end{tabular}
\end{center}
\caption{Consolidated investor balance sheet in period $t$.}
\label{tab:nonbankbs}
\end{table}%

Throughout this paper, we restrict attention to initial balance sheets that are non-degenerate in the sense that both banks and investors hold some bonds ($\bankBonds{0}, \investorBonds{0} > 0$) and the banks are solvent ($\bankEquity{0} \geq 0$).

\subsection{Balance sheet consistency}
Critical to our analysis is the concept of \emph{balance sheet consistency}, meaning a pair of portfolios and a price of bonds such that banks and investors are both maximizing their respective objective functions, given their shared belief about the bonds' future value. Initially, the banks and investors have balance sheets (i.e., Tables~\ref{tab:bankbs} and \ref{tab:nonbankbs} at time $t=0$) that are consistent with respect to a common belief about the bonds' value. This shared belief changes $T$ many times via the arrival of unanticipated news that is good or bad. At each time step $t \in \{1,2, \dots, T\}$, the banks and investors restore consistency of their balance sheets with respect to the new belief at time $t$ by trading, which results in a new set of portfolios. This trade also changes the bond price and the {money supply}, defined as the sum of deposits and currency in circulation (i.e., $\deposits{t}+\currency{t}$). Uncertainty regarding the bonds' value becomes resolved in period $T+1$.

To define consistency formally, suppose that banks were to buy $d$ units of bonds from investors at the price $\price{0}$. This purchase would require the banks to transfer $\price{0} d$ units of money to the investors by giving them $\currencyFraction \price{0} d$ units of currency (drawn from the banks' reserves) and by crediting the investors' deposit accounts by $(1-\currencyFraction) \price{0} d$ in the aggregate.\footnote{An investor who sells bonds to one bank may deposit the proceeds in another bank, but here we model these changes at the consolidated level of the banking sector, ignoring identities of banks.} Table~\ref{tab:aftertrade} shows the resulting changes to balance sheets.

\begin{table}[htdp]
\begin{center}
\begin{tabular}{l l l}
Bank assets   & Bank liabilities  & Investor assets  \\ \hline
bonds $ +\price{0}d$& deposits $+ (1-\currencyFraction) \price{0} d$
 & bonds $-\price{0} d$   \\
reserves $- \currencyFraction \price{0} d$ &  
&  currency $+\currencyFraction \price{0} d$  \\ 
 & 
 & deposits $+ (1-\currencyFraction) \price{0} d$
\end{tabular}
\end{center}
\caption{Balance sheet changes following a purchase of $d$ bonds by the banks at price $\price{0}$ in period $t=0$. 
}
\label{tab:aftertrade}
\end{table}%

Note that if the banks were to purchase bonds (i.e., if $d >0$), then the money supply (the sum of deposits and currency in circulation) would rise. Conversely, if the banks were to sell bonds (i.e., if $d<0$), then the money supply would contract. 

At each period $t \in \{0, 1, \dots, T\}$, we assume that the banks and investors share a common belief $V_t$ about what each bond will be worth when uncertainty is resolved in period $T+1$. Each belief $V_t$ is a continuous random variable with density $f_t$ that has support $[0,1]$ and expected value $\E V_t$. The banks maximize their expected equity subject to the constraint that their probability of insolvency (when uncertainty is resolved) is smaller than $\epsilon$, a fixed parameter in $(0,1)$.\footnote{A popular instantiation of this type of insolvency constraint is known as ``Value-at-Risk'' or VaR~\citep{Shin2010_RiskandLiquidity}. We note, however, that VaR plays a dual role in the banking system: first, as a tool for risk assessment, where it measures balance sheet strength by estimating the loss in value such that losses greater than it occur with probability no larger than a fixed, small number such as $1\%$; and second, as a strategy for risk management [e.g., setting VaR equal to equity limits the perceived likelihood of insolvency~\citep{Longworth2010}]. Here we invoke VaR in this latter sense of risk management.} 
Meanwhile, the investors maximize their expected utility and are risk averse, with preferences given by a strictly increasing, strictly concave function of terminal wealth $u : (0, \infty) \to \mathbb{R}$. 

Now we define balance sheet consistency. 
\begin{definition}
The variables $(\price{t}, \currency{t}, 
\reserves{t}, \bankBonds{t}, \investorBonds{t}, \deposits{t})$ are \emph{consistent} with respect to the common belief $V_t$ if and only if \begin{enumerate} \item the banks are maximizing their expected equity subject to their insolvency constraint, i.e.,
\begin{align*}
0 = & \argmax_{-\bankBonds{t} \, \leq \, d \, \leq  \, \reserves{t} / (\currencyFraction \price{t})}  \E \left [ V_t (\bankBonds{t} + d)+\reserves{t}-(\deposits{t} + \price{t} d) \right ] \\ & \text{ subject to } \prob{\text{insolvent}} \leq \epsilon, 
\end{align*}
where $\{\text{insolvent}\} = \{V_t(\bankBonds{t}+d) + \reserves{t} - (\deposits{t} + \price{t} d) < 0\}$; and
\item the investors are maximizing their expected utility
\begin{align*}
0 = \argmax_{-\investorBonds{t} \, \leq \, d \, \leq \, \deposits{t} / [\price{t} (1-\currencyFraction)]} \E u \left [ V_t (\investorBonds{t} - d) + \deposits{t} + c_t + \price{t} d \right ].
\end{align*}
\end{enumerate}
\label{def:consistency}
\end{definition}

The constraints on $d$ in Def.~\ref{def:consistency}, as discussed below, simply ensure that the banks and investors do not sell more bonds than they hold nor buy more bonds than they can finance. The following numerical example, to which we will refer throughout our analysis, illustrates the idea of consistent balance sheets. 

\begin{example} 
\label{ex:consistent}
The balance sheets in Table~\ref{tab:consistent_example} are consistent with respect to the belief $V_0 \sim \text{Beta}(20,2)$ if $\epsilon = 1\%$ $($i.e., the banks ensure that insolvency occurs with probability at most $\epsilon = 1\%)$; the investors hold $\currencyFraction = 10\%$ of their money as currency; and the investors maximize the utility function $u(w) = w^{1-\lambda}/(1-\lambda)$ with $\lambda = 15$ $($i.e., investors' preferences satisfy constant relative risk aversion$)$. The bond price is $\price{0} = \$0.87$ per unit of face value.

\begin{table}[htdp]
\begin{center}
\begin{tabular}{l l l l}
Bank assets & Bank liabilities & Investor assets & Investor liabilities \\ \hline
$439$ bonds at price $ \$ 0.87$ & deposits $\$486$
 & $611$ bonds at $\$ 0.87$ & equity $\$1072$ \\
reserves $\$168$ & equity $\$62$
&  currency $\$ 54$  \\ 
& & deposits $\$486$
\end{tabular}
\end{center}
\caption{Consistent balance sheets for the beliefs and preferences in Example~\ref{ex:consistent}.}
\label{tab:consistent_example}
\end{table}%
\end{example}

Next we examine how changes in beliefs affect balance sheets and the market-clearing bond price.

\subsection{Belief Shocks}
\label{sec:demandFunctions}

Consider a financial system with a pair of consistent balance sheets at time $t \in \{0, 1, \dots, T-1\}$, and suppose that new information about the bonds' value emerges in period $t+1$ (e.g., news of higher than expected foreclosures in the subprime mortgage market). We assume that this information is public and interpreted identically by the banks and by the investors. Nevertheless, because the banks and investors have different objective functions, the news may cause them to trade and hence reach a new market-clearing price. We consider shocks to the belief $V_t$ that  shift probabilities to higher values (positive shocks) or to lower values (negative shocks).

\begin{definition}
$V_{t+1}$ is a \emph{positive shock} to $V_t$ (or equivalently $V_t$ is a \emph{negative shock} to $V_{t+1}$) if and only if 
\begin{enumerate} \item $V_{t+1}$ second-order stochastically dominates\footnote{Second-order stochastic dominance means that $\int_0^x F_{t+1}(x) dx > \int_0^x F_{t}(x) dx$ for all $x \in [0,1]$, where $F_t(x)$ is the CDF of $V_t$.} $V_t$, and \item $v_{t+1} > v_t$, where $v_t$ is the first $\epsilon$-quantile\footnote{The first $\epsilon$-quantile of $V_t$, denoted $v_t$, is the inverse of the CDF of $V_t$ evaluated at $\epsilon$, $F_t^{-1}(\epsilon)$.} of $V_t$. \end{enumerate}
\end{definition}

To determine the changes in the portfolios and the change in the bond price that would restore consistency after a shock, we express the investors' demand for bonds $\investorDemand{\price{}}$ and the banks' demand for bonds $\bankDemand{\price{}}$ in terms of a hypothetical price $\price{} \in [0,1]$, given the consistent balance sheets in period $t$. A market-clearing price must make excess demand vanish in the aggregate, so the new market-clearing price $\price{t+1}$ is a root of the total demand $\totalDemand{\price{}} \equiv \investorDemand{\price{}}+ \bankDemand{\price{}}$. The banks' and investors' demands at this price determine their new portfolios.

\subsection{Investor Demand}
As in the definition of consistency (Definition~\ref{def:consistency}), after the belief $V_t$ is replaced by $V_{t+1}$, the investors adjust portfolios to maximize the expected value of their utility. Because the proportion of the investors' money in currency before and after the purchase of bonds is $\currencyFraction$, the investors can buy $d$ units of bonds using $\currencyFraction \price{} d$ dollars in cash and $(1 - \currencyFraction) \price{} d$ dollars from their deposit accounts. Investors cannot borrow, so their total demand is upper-bounded by their holdings of money (currency plus deposits). That is, investors can demand at most $d \leq \deposits{t} / [ \price{} (1-\currencyFraction) ] = \currency{t}/ (\price{} \currencyFraction)$ many bonds. Also, the investors cannot sell more bonds than they hold, so they must demand at least $d \geq -\investorBonds{t}$. Thus the investors demand 
\begin{align}
 \investorDemand{\price{}; \currency{t}, \deposits{t}, \investorBonds{t}} 
&:= \argmax_{-\investorBonds{t} \, \leq \, d \, \leq \, \deposits{t}/[\price{} (1-\currencyFraction)]}\E \left ( u \left [ V_{t+1}(\investorBonds{t}+d) + \deposits{t} + \currency{t} - \price{} d \right ] \right ) \label{eq:nonbankdemand} \\
&= \argmax_{-\investorBonds{t} \, \leq \, d \, \leq \, \deposits{t}/[\price{} (1-\currencyFraction)]} \int_0^1 u \left [ v(\investorBonds{t}+d) + \deposits{t} + \currency{t} - \price{} d \right ] f_{t+1}(v) dv, \notag
\end{align}
where $f_{t+1}(\cdot)$ is the probability density of $V_{t+1}$. Lemma~\ref{lem:Dn} in~\ref{sec:investordemand} shows that, under mild assumptions satisfied here, the investors' demand~\eqref{eq:nonbankdemand} is single-valued at all prices (and hence can be expressed as a function). 

Following Example 1, we consider the special case with investor utility $u(\cdot)$ that satisfies constant relative risk aversion (CRRA) and Beta-distributed beliefs. That is, the investors' utility function $u(w; \lambda) := w^{1-\lambda}/(1-\lambda)$, where $\lambda > 0$ and $\lambda \neq 1$, and the beliefs of bond values at time $t \in \{0, 1, 2, \dots, T\}$ are
$V_{t} \sim \text{Beta}(\alpha_{t}, \beta_{t})$ for some $\alpha_{t}, \beta_{t} > 0$. 

\begin{figure}[t]
\begin{center}
\includegraphics{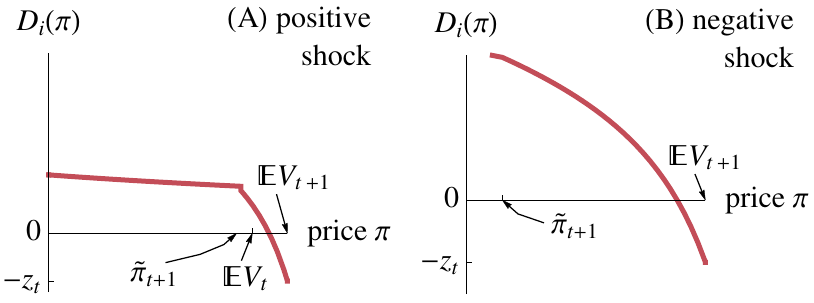}
\caption{The investor demand function $\investorDemand{\price{}}$ [Eq.~\eqref{eq:nonbankdemand}] is non-increasing in the price $\price{}$. By Lemma~\ref{lem:Dn}, we know that the investors demand $\investorDemand{\price{}} = -\investorBonds{t}$ if and only if $\price{} \geq \E V_{t+1}$. Here, investor preferences are represented by the CRRA utility function $u(w) := w^{1-\lambda}/(1-\lambda)$ for $0<\lambda \neq 1$; for this case, we know by Lemma~\ref{lem:CRRA} in~\ref{sec:investordemand} that the investors buy as much as they can afford, $\investorDemand{\price{}} = \deposits{t} / [\price{} (1-\currencyFraction)]$, if and only if $\pi \leq \priceDemandLessThanMax_{t+1}$.
\label{fig:nonbankdemand}}
\end{center}
\end{figure}

Figure~\ref{fig:nonbankdemand} illustrates that in this case the investor demand function $\investorDemand{\price{}}$ is non-increasing with the price $\price{}$, and there is some price strictly below $\E V_{t+1}$ above which demand becomes negative. Lemma~\ref{lem:CRRA} in~\ref{sec:investordemand} shows that the investors' demand falls below $\deposits{t+1}/[\price{}(1-\currencyFraction)]$, the maximum number of bonds that the investors can afford, precisely at the price $\priceDemandLessThanMax_{t+1} \equiv (\alpha_{t+1} - \lambda)/(\alpha_{t+1} + \beta_{t+1} - \lambda)$.

\subsection{Bank Demand}
\label{sec:bank_demand}
Recall that the banks maximize their expected equity subject to an insolvency constraint. Consider a hypothetical price $\price{} \in [0,1]$ after the shock that replaces the belief $V_t$ by $V_{t+1}$. If this hypothetical price $\price{}$ would render the banks insolvent (i.e., if $\price{} \bankBonds{t} + \reserves{t} - \deposits{t} < 0$), then the banks are forced to sell all their bonds, so they demand $\bankDemand{\price{}} := -\bankBonds{t}$. Otherwise, the banks remain solvent, so they demand a number of bonds $d$ that maximizes their expected equity in period $T+1$ using the new belief $V_{t+1}$,
\begin{align}
e_{t+1} = V_{t+1} (\bankBonds{t}+d) + \reserves{t} - (\deposits{t}+\price{} d),
\label{eq:bank_equity}
\end{align}
subject to the three constraints 
\begin{subequations}
\begin{align}
 d  \geq -\bankBonds{t}  \quad \text{(can sell at most all their bonds),}&
\label{eq:dMin} \\
  d \currencyFraction \price{} \leq \reserves{t}  \quad \text{(cannot have negative reserves),}\label{eq:dMax} \\
 \prob{\bankEquity{t+1}<0}  \leq \epsilon  \quad \text{(insolvency constraint).}& \label{eq:VaRconstraint}
\end{align}
\label{eq:bankDemandConstraints}
\end{subequations}
Thus, the banks' demand function is
\begin{align}
&\bankDemand{\price{}; \reserves{t}, \bankBonds{t}, \deposits{t}} = 
\begin{dcases} 
\argmax_{\substack{d \text{ s.t.~\eqref{eq:bankDemandConstraints}}}}  \E \bankEquity{t+1} &  \parbox{7.6em}{ if $\price{} \bankBonds{t} +\reserves{t}  - \deposits{t}\geq 0$ } \\ 
-\bankBonds{t} &  \text{else}
\end{dcases}.
\label{eq:bankdemand_VaR_definition}
\end{align}

Next we simplify the insolvency constraint~\eqref{eq:VaRconstraint}. Define $\ve{\epsilon}{t+1}$ as the first $\epsilon$-quantile of the belief $V_{t+1}$. That is, under belief $V_{t+1}$, the probability that the terminal value of bonds will be at least $\ve{\epsilon}{t+1}$ is $1 - \epsilon$. Because $V_{t+1}$ is continuous, the insolvency constraint~\eqref{eq:VaRconstraint} is equivalent to
\begin{align}
\ve{\epsilon}{t+1} (\bankBonds{t} + d) + \reserves{t} - (\deposits{t} + \price{} d) \geq 0.
\label{eq:solvencyequivalence}
\end{align}
From~\eqref{eq:solvencyequivalence}, we see that the banks poise themselves at $\epsilon$-probability of insolvency 
[i.e., $\prob{\text{insolvent}} = \epsilon$] by demanding
\begin{align}
d = \frac{ \ve{\epsilon}{t+1} \bankBonds{t} + \reserves{t} - \deposits{t}}{\price{} -  \ve{\epsilon}{t+1}}
\label{eq:poise_at_VaR}
\end{align}
many bonds, provided that $\price{} > \ve{\epsilon}{t+1}$. 

\begin{figure}[t]
\begin{center}
\includegraphics{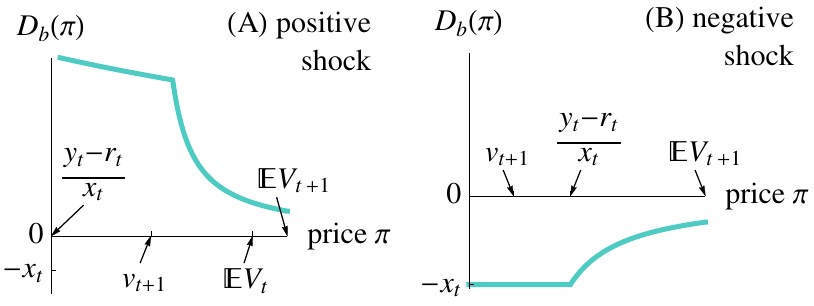}
\caption{Illustrations of the banks' demand function, Eq.~\eqref{eq:bankdemand_VaR_definition}, for positive and negative shocks. Immediately after a positive shock [panel (A)], the banks comply with their insolvency constraint~\eqref{eq:solvencyequivalence} (i.e., $\ve{\epsilon}{t+1} \bankBonds{t} + \reserves{t} - \deposits{t} >0$); in this case, the banks want to buy bonds provided that the price $\price{} \leq \E V_{t+1}$ and provided that the banks are currently solvent ($\price{} \bankBonds{t} + \reserves{t} - \deposits{t} \geq 0$); the banks' demand is given by Eq.~\eqref{eq:bankDemand:positiveShock}. Immediately after a negative shock [panel (B)], the banks violate their insolvency constraint~\eqref{eq:solvencyequivalence} (i.e., $\ve{\epsilon}{t+1} \bankBonds{t} + \reserves{t} - \deposits{t} < 0$); in this case, the banks want to sell bonds, and the banks' demand is given by Eq.~\eqref{eq:bankDemand:negativeShock}. 
The plot ranges are identical in Figs.\ \ref{fig:nonbankdemand}(A) and\ \ref{fig:bankdemand}(A) and in Figs.\ \ref{fig:nonbankdemand}(B) and\ \ref{fig:bankdemand}(B); the sums of these demands 
are shown in Fig.~\ref{fig:shock_equations} of Sec.~\ref{sec:illustrate_total_demand}. 
\label{fig:bankdemand}}
\end{center}
\end{figure}

In~\ref{sec:derive_bank_demand}, 
we combine Eq.\ \eqref{eq:poise_at_VaR} with the constraints~\eqref{eq:dMin} and \eqref{eq:dMax}; here we give the result. If $\ve{\epsilon}{t+1} > \ve{\epsilon}{t}$, as occurs in a positive shock, then the numerator $\ve{\epsilon}{t+1} \bankBonds{t} + \reserves{t} - \deposits{t}$ of Eq.\ \eqref{eq:poise_at_VaR} is positive, so the banks satisfy their insolvency constraint~\eqref{eq:solvencyequivalence} after the shock. Then the banks want to buy bonds according to the decreasing function
\begin{subequations}
\begin{align}
\bankDemand{\price{}} &= 
\begin{dcases}
     \frac{\reserves{t}}{\currencyFraction  \price{}} & \text{if } 
     \price{} < \frac{\ve{\epsilon}{t+1}}{1-\frac{\currencyFraction}{\reserves{t}} (\ve{\epsilon}{t+1} \bankBonds{t} + \reserves{t} - \deposits{t})} \\
	\frac{ \ve{\epsilon}{t+1} \bankBonds{t} + \reserves{t} - \deposits{t}}{\price{} -  \ve{\epsilon}{t+1}}  & \text{else} 
\end{dcases}
\label{eq:bankDemand:positiveShock}
\end{align}
as long as the banks are currently solvent ($\price{} \bankBonds{t} + \reserves{t} - \deposits{t} \geq 0$) and the price $\price{} \le \E V_{t+1}$. 

On the other hand, if $\ve{\epsilon}{t+1} < \ve{\epsilon}{t}$, as occurs in a negative shock, then the numerator $\ve{\epsilon}{t+1} \bankBonds{t} + \reserves{t} - \deposits{t}$ of Eq.~\eqref{eq:poise_at_VaR} is negative, meaning that the banks violate their insolvency constraint~\eqref{eq:solvencyequivalence} immediately after the shock. Then the banks will sell bonds at all prices by demanding
\begin{align}
\bankDemand{\price{}} &= 
\frac{ \ve{\epsilon}{t+1} \bankBonds{t} + \reserves{t} - \deposits{t}}{\price{} -  \ve{\epsilon}{t+1}}
\label{eq:bankDemand:negativeShock}
\end{align}
\label{eq:bankDemand_shocks}
\end{subequations}
as long as the banks are currently solvent (i.e., $\price{} \bankBonds{t} + \reserves{t} - \deposits{t} \geq 0$) and the price $\price{} \le \E V_{t+1}$. Notice in Eq.\ \eqref{eq:bankDemand_shocks} and in Fig.\ \ref{fig:bankdemand}(B) that for negative shocks the banks' demand \emph{increases} with the price $\price{}$. In contrast to typical demand functions, the banks sell \emph{more aggressively} at \emph{lower} prices. This increasing demand function results in a non-monotonicity of aggregate demand in the face of negative shocks. 

\subsection{Aggregate Demand}
\label{sec:illustrate_total_demand}

\begin{figure*}[t]
\begin{center}
\includegraphics[width=\textwidth]{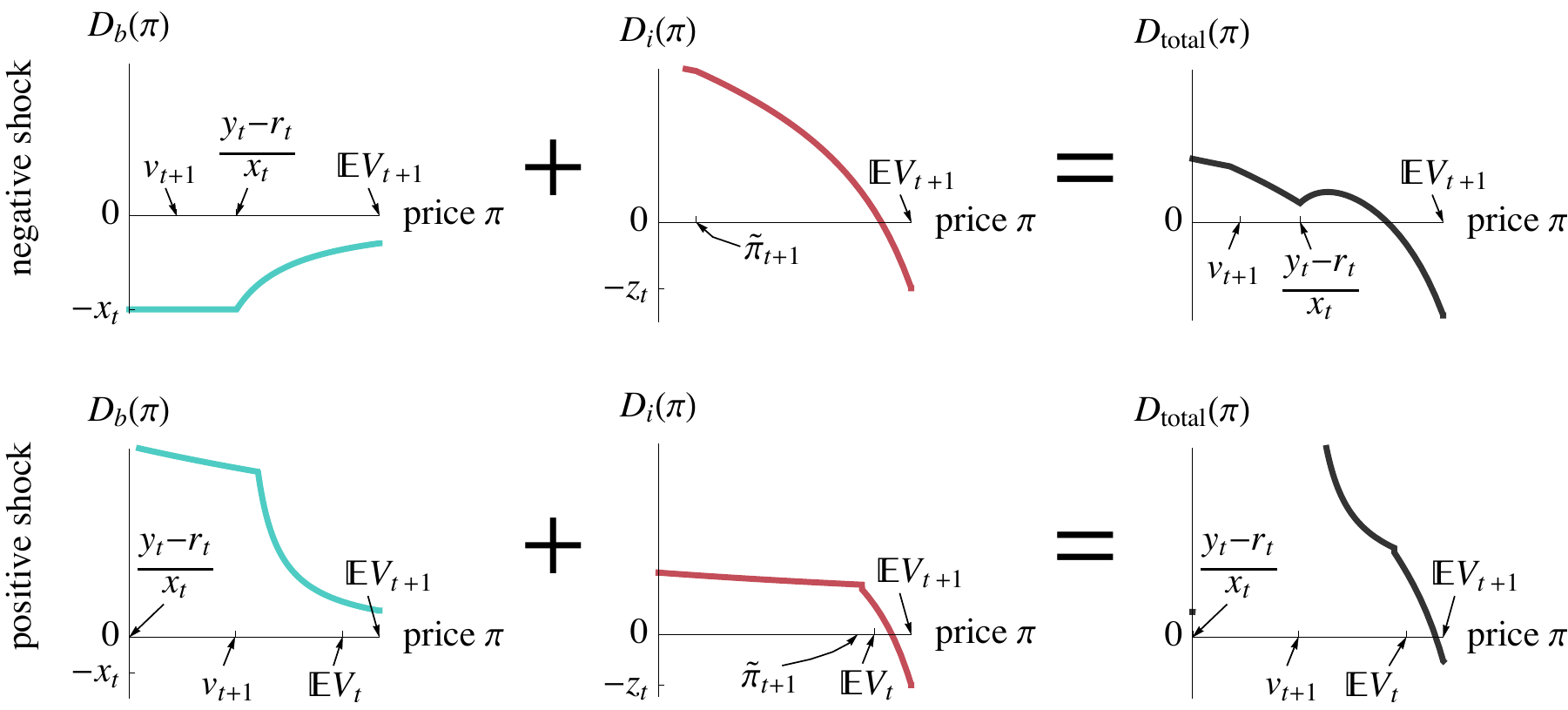} 
\caption{Illustrations of the bank demand function $\bankDemand{\price{}}$ (left column), the investor demand function $\investorDemand{\price{}}$ (middle column), and their sum, the total demand function $\totalDemand{\price{}}$ (right column), for a negative shock (top row) and for a positive shock (bottom row). Within each row, the plot ranges are identical. The plots in the left and middle columns are identical to Figs.~\ref{fig:nonbankdemand}--\ref{fig:bankdemand}.}
\label{fig:shock_equations}
\end{center}
\end{figure*}

In Fig.~\ref{fig:shock_equations}, we plot the total demand function $\totalDemand{\price{}} \equiv \bankDemand{\price{}} + \investorDemand{\price{}}$, the roots of which are the new equilibrium prices. 
The plot ranges in the top and bottom rows of Fig.~\ref{fig:shock_equations} are identical. Notice that the banks' demand in the case of a negative shock (top-left plot of Fig.~\ref{fig:shock_equations}) is increasing in the bond price, which can make the total demand (top-right plot of Fig.~\ref{fig:shock_equations}) non-monotonic and hence have multiple roots (i.e., multiple equilibrium prices $\price{t+1}$). Next we show that at one of those multiple equilibria the banks are insolvent and that banking crises can appear ``out of the blue''.

\section{Results}
\label{sec:results}
\subsection{Negative Shocks}
\label{sec:shocks:negative}
\begin{figure}[!t]
\begin{center}
\includegraphics[trim = 0 0 0 1]{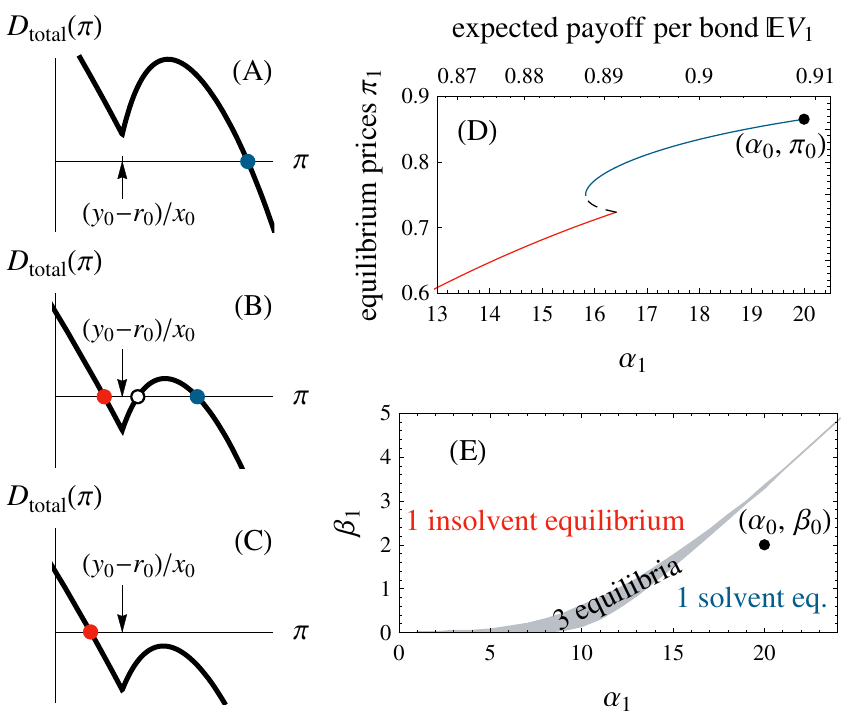}
\caption{Cusp catastrophe from a negative shock. The initial condition is Example~\ref{ex:consistent}, in which the initial belief $V_0 \sim \text{Beta}(\alpha_0, \beta_0)$ with $(\alpha_0,\beta_0) = (20,2)$. In panels (A)--(D), we consider negative shocks $V_1 \sim \text{Beta}(\alpha_1, \beta_1)$ for progressively smaller $\alpha_1$, with $\beta_1$ fixed at $\beta_0 = 2$. 
Panels (A)--(C) show the total demand function $\totalDemand{\price{}}$, the roots of which are the market-clearing prices $\price{1}$; filled and empty circles in (A)--(C) denote stable and unstable equilibria, respectively; the banks are solvent at an equilibrium price $\price{1}$ if and only if $\price{1} \geq (\deposits{0} - \reserves{0})/\bankBonds{0}$. The following behaviors occur as $\alpha_1$ decreases: the negative shock becomes more severe because the expected payoff from bonds decreases according to $\E V_1 = \alpha_1/(\alpha_1+\beta_1)$ [see the top axis of (D)]; the largest equilibrium price [the right-most root $\price{1}$ in (A)--(C), the maximum of the curves in (D)] decreases, at first slowly and then precipitously when $\alpha_1$ passes the region of three equilibria [compare panels (B) and (C) and notice the region of three equilibria near $\alpha_1 =16$ in (D)]; after passing the region of three equilibria, there exists only one equilibrium, at which the banks are insolvent [red dot in (C), red curve in (D)]. The gray region in (E) marks the belief $(\alpha_1, \beta_1)$ that gives rise to three equilibria [as in (B)]; above that region, the price drops sharply, and the banks are insolvent, i.e., $\price{1} \bankBonds{1} + \reserves{1} - \deposits{1} < 0$ [see (C)]. 
\label{fig:negativeshock}}
\end{center}
\end{figure}

Starting from a pair of consistent balance sheets, a small negative shock to beliefs would lead to a slightly smaller bond price. We illustrate that market-clearing price using a blue dot in Fig.\ \ref{fig:negativeshock}(A). A slightly stronger negative shock could create two new equilibrium prices, marked by a red dot and by a black circle in Fig.\ \ref{fig:negativeshock}(B). Those two new prices appear in a saddle-node bifurcation at the price $(\deposits{t} - \reserves{t})/\bankBonds{t}$. An even stronger negative shock annihilates the two larger equilibrium prices in another saddle-node bifurcation [Fig.\ \ref{fig:negativeshock}(C)]. Stable and unstable roots of $\totalDemand{\price{}}$ are depicted by filled and empty circles, respectively, in Fig.\ \ref{fig:negativeshock}.

The price $(\deposits{t} - \reserves{t})/\bankBonds{t}$ marks not only where two new equilibrium prices appear but also whether the banks become insolvent, because the banks are solvent at a hypothetical price $\price{}$ if and only if $\price{} \bankBonds{t} + \reserves{t} - \deposits{t} >0$. The prices marked by red dots in Figs.\ \ref{fig:negativeshock}(B)--(C) and by the red line in Fig.\ \ref{fig:negativeshock}(D) are equilibrium prices that result in an insolvent banking sector [$\price{} <(\deposits{t} - \reserves{t})/\bankBonds{t}$], whereas the blue dots in Figs.\ \ref{fig:negativeshock}(A)--(C) and the blue line in Fig.\ \ref{fig:negativeshock}(D) denote prices that maintain solvency.

Figures\ \ref{fig:negativeshock}(A) and\ \ref{fig:negativeshock}(B) demonstrate that a small shock can create an equilibrium at which the banking sector is insolvent [i.e., create the red dot in Fig.\ \ref{fig:negativeshock}(B)] even if there continues to exist an equilibrium price close to the original price and at which the banks are solvent [i.e., the blue dot exists in Fig.\ \ref{fig:negativeshock}(B)]. This multiplicity of equilibria implies that a widespread fear of a crisis can become self-fulfilling and result in a collapse of asset prices, despite the fact that no crisis would occur in the absence of such fear. Actions by the central bank in supporting asset prices and in containing fear can effectively coordinate expectations on the more optimistic equilibrium [the blue dot in Fig.~\ref{fig:negativeshock}(B)]. However, if the shock is sufficiently large, as in Fig.\ \ref{fig:negativeshock}(C), then coordinating expectations alone cannot prevent crisis because there no longer exists an equilibrium with a solvent banking sector. 

In summary, negative shocks lead to a cusp catastrophe~\citep{Rosser2000}. Specifically, there is a region of shocked beliefs $(\alpha_1, \beta_1)$, marked in gray in Fig.\ \ref{fig:negativeshock}(E), that give rise to three equilibrium prices. Uniqueness of equilibrium is restored for beliefs above that gray region, but this equilibrium is a very different kind because it entails widespread bank insolvency. Next, we show that positive shocks increase the risk of such a crisis.

\subsection{Positive Shocks}
\label{sec:shocks:positive}
Now suppose that beliefs about the terminal value of the bonds becomes more optimistic because, for example, the prices of underlying assets (e.g., housing prices) rise faster than expected. This positive shock generates slack in the insolvency constraint~\eqref{eq:VaRconstraint}, which allows the banks to undertake more risk by buying bonds and by expanding their balance sheets. Because the investors share the more optimistic beliefs, they also wish to buy bonds at the price that prevailed before the arrival of the positive shock. Thus, there is excess demand for bonds at the initial price, so the price will accordingly rise\footnote{Recall, our model assumes that the total volume of bonds is fixed, thus ruling out the possibility that banks can increase the size of their balance sheets by underwriting new assets.}. 

The price will rise until the investors are enticed to sell bonds to the banks. 
To see why, recall that after a positive shock the banks demand to buy bonds at all prices below the new expected value $\E V_1$ [recall Fig.\ \ref{fig:bankdemand}(A)]. The investors, meanwhile, demand to sell bonds if and only if the price becomes large enough [recall from Fig.\ \ref{fig:bankdemand}(A) that their demand becomes negative just below $\E V_1$]. Thus, the new market-clearing price will be determined by the investors selling bonds to the banks. The banks correspondingly expand their balance sheets and increase the money supply as they credit the investors' deposit accounts. 
 
\begin{figure}[t]
\centering
\includegraphics[trim = 0 0 0 0]{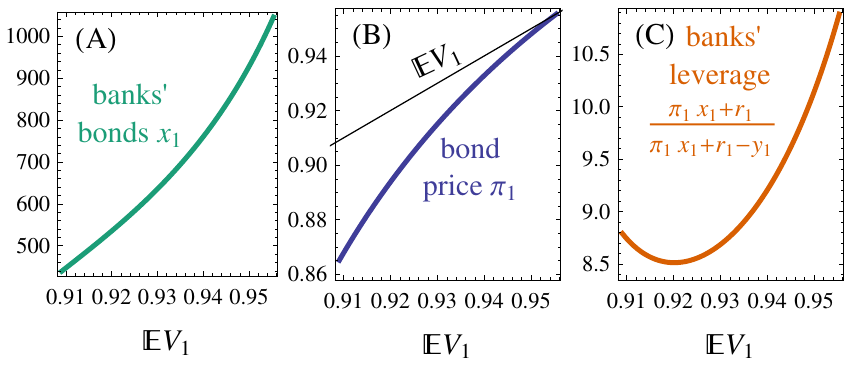}
\caption{
A positive belief shock induces the banks to buy bonds (A), to bid up the bond price (B), and typically to increase their leverage (C). Here, we subject Example~\ref{ex:consistent} to a belief shock $V_1 \sim \text{Beta}(\alpha_1, \beta_1)$, for $\alpha_1 \in (20,42.8]$ and $\beta_1 = \beta_0 = 2$. 
The  axis $\E V_1 = \alpha_1 / (\alpha_1 + \beta_1)$ is the severity of the positive shock. 
\label{fig:shocks:positive}}
\end{figure}

As the positive shock becomes stronger (larger expected bond value $\E V_1$), the banks buy more bonds [Fig.\ \ref{fig:shocks:positive}(A)] at an ever higher price $\price{1}$ [Fig.\ \ref{fig:shocks:positive}(B)]. Bank leverage (the ratio of assets to equity) rises if the news is sufficiently positive [Fig.\ \ref{fig:shocks:positive}(C)]. Next, we show that this procyclical leverage worsens the risk of bankruptcy when beliefs subsequently become pessimistic.

\subsection{Reversal of a Positive Shock}
\label{sec:shocks:reversal}

Consider a positive shock followed by a negative shock that restores the initial belief. One might expect that such shocks of equal magnitude ``cancel'' and leave the financial system unchanged. However, this model is nonlinear, so portfolios depend on the history of beliefs. Moreover, a sufficiently strong positive shock [that results in greater bank leverage; see Fig.\ \ref{fig:shocks:positive}(C)] that is followed by an equally strong negative shock can result in insolvency.

\begin{figure}[t]
\begin{center}
\includegraphics[]{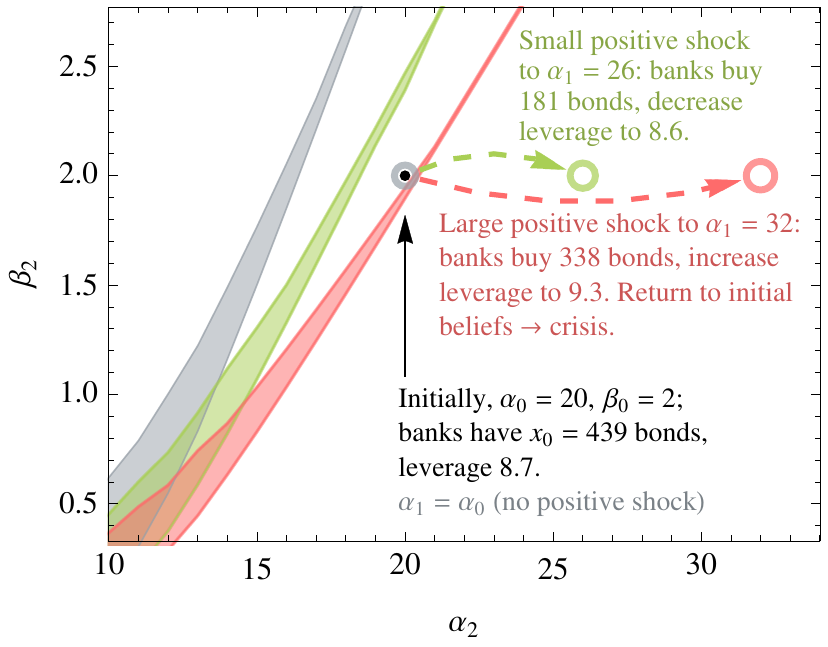}
\caption{A sufficiently strong positive belief shock followed by an equally negative shock can result in insolvency. Here, the initial condition is given by Example~\ref{ex:consistent}, in which the initial belief $V_0 \sim \text{Beta}(\alpha_0, \beta_0)$ with $(\alpha_0,\beta_0) = (20,2)$, marked by a black dot. At time $t=1$, a positive shock changes the belief to have parameters $\beta_1 = \beta_0$ and $\alpha_1 = 20, 26, 32$ (gray, green, and red circles, respectively). The horizontal and vertical axes are the parameters $(\alpha_2, \beta_2)$ of the new belief at time $t=2$. The three colored regions denote the belief parameters $(\alpha_2,\beta_2)$ such that there exist three equilibrium prices $\price{2}$. The gray region is identical to that in Fig.\ \ref{fig:negativeshock}(E). As in Fig.~\ref{fig:negativeshock}(E), above the colored region, the banks are insolvent at the unique equilibrium price. Note that the black dot lies above the red region, meaning that restoring the initial beliefs after a large positive shock to $\alpha_1 = 32$ (red circle) leads to insolvency.}
\label{fig:positive_then_negative}
\end{center}
\end{figure}

Figure~\ref{fig:positive_then_negative} illustrates such a crisis caused by optimism that is reversed. The initial belief $(\alpha_0, \beta_0)$ is marked by a black dot. At step $1$, the belief undergoes either a small positive shock, a large positive shock, or no shock at all; the resulting belief parameters $(\alpha_1, \beta_1)$ are marked in Fig.~\ref{fig:positive_then_negative} by the green circle, red circle, and gray circle, respectively. Next, consider in step $2$ a second shock to a belief $V_2 \sim \text{Beta}(\alpha_2, \beta_2)$. The colored regions in Fig.~\ref{fig:positive_then_negative} are the regions of beliefs $(\alpha_2, \beta_2)$ that give rise to three equilibrium prices [just like Fig.~\ref{fig:negativeshock}(E)].

We focus in particular on what happens if the initial belief is restored in period $2$, i.e., if $(\alpha_2, \beta_2)$ equals $(\alpha_0, \beta_0)$, the black dot. In this case, the initial and final beliefs are identical; we only vary the intermediate belief in period $1$. This intermediate state crucially moves the region of multiple equilibria for period $2$. For the small positive shock (the green circle and green region in Fig.\ \ref{fig:positive_then_negative}), the initial belief (black dot) lies to the right of the green region; thus, restoring the initial belief after the small positive shock does not threaten the banks' solvency. However, for the large positive shock (the red circle and red region), the initial belief lies to the \emph{left} of the red region, so restoring the initial belief after the larger positive shock moves the financial system across the region of multiple equilibria and into the region of certain insolvency. To reiterate, reversing a large positive shock can precipitate widespread bank insolvency. 
This behavior is consistent with a number of historical episodes, including the global financial crisis of 2007--2009, in which a rise and subsequent fall in real estate prices led to the near collapse of the banking system.

Up to this point we have assumed that banks are unconstrained by regulatory requirements. Next we consider the effects of capital requirements. 

\section{Capital Requirements}

A capital requirement aims to prevent insolvency crises by requiring that banks maintain sufficient levels of capital (equity) relative to their risk-weighted assets. In accordance with the standards in Basel III, we assign risk weight zero to bank reserves and risk weight one to the banks' bond holdings. With this assumption, the banks' \emph{capital-to-assets ratio} $\gamma_t$ is the banks' equity $\bankEquity{t}$ divided by the total, market-value of their bond holdings, $\price{t} \bankBonds{t}$. Then the \emph{capital requirement} mandates that this capital-to-assets ratio $\gamma_t$ equals or exceeds some minimum amount $\gamma_t^\text{min} \in [0,1]$, 
\begin{align}
\gamma_t \equiv \frac{\bankEquity{t}}{\price{t} \bankBonds{t}} \equiv \frac{\price{t}\bankBonds{t} + \reserves{t} - \deposits{t}}{\price{t} \bankBonds{t}} \geq \gamma_t^\text{min}.
\label{eq:CARrule}
\end{align}
at each step $t \in \{0, 1, \dots, T\}$

Next we examine the effect of this constraint on the banks' demand in the events of positive and negative shocks.  

\subsection{Capital Requirements and Bank Demand}

If the banks were to buy $d$ bonds at a new price $\price{}$ in period $t+1$, then the balance sheets would become those in Table~\ref{tab:aftertrade}, as discussed above. The capital requirement~\eqref{eq:CARrule} gives another constraint on the banks' demand $d$,
\begin{align*}
\gamma_{t+1} = \frac{\price{}(\bankBonds{t} + d)+ \reserves{t} - \currencyFraction \price{} d - \left [ \deposits{t} + (1-\currencyFraction) \price{} d \right ]}{\price{} (\bankBonds{t} + d)} \geq \gamma_{t+1}^\text{min},
\end{align*}
or, upon rearranging, 
\begin{align}
d &\leq 
-\bankBonds{t} + \frac{\price{} \bankBonds{t} + \reserves{t} - \deposits{t}}{\price{} \gamma_{t+1}^\text{min}} 
= \bankBonds{t} \left (\frac{1}{\gamma_{t+1}^\text{min}} -1 \right ) + \frac{\reserves{t} - \deposits{t}}{\price{} \gamma_{t+1}^\text{min}}.
\label{eq:CAR:upperbound}
\end{align}
In the event that this upper bound~\eqref{eq:CAR:upperbound} on the banks' demand $d$ is smaller than $-\bankBonds{t}$, we define the capital constraint to be $d = -\bankBonds{t}$ because the banks cannot sell more bonds than they hold. Thus, the cap on bank demand due to the capital requirement, denoted by $\CARconstraint{\price{}}$, is defined to be 
\begin{align}
\CARconstraint{\price{}; \reserves{t}, \bankBonds{t}, \deposits{t}, \gamma_{t+1}^\text{min}}  
\equiv 
\max \left \{
-\bankBonds{t}, 
-\bankBonds{t} + \frac{\price{} \bankBonds{t} + \reserves{t} - \deposits{t}}{\price{} \gamma_{t+1}^\text{min}}
\right \}.
\label{eq:CAR:definition}
\end{align}
Augmenting the banks' demand function~\eqref{eq:bankdemand_VaR_definition} with the capital constraint $d \leq \CARconstraint{\price{}}$ results in the demand function
\begin{align}
\bankDemand{\price{}; \reserves{t}, \bankBonds{t}, \deposits{t}, \gamma_{t+1}^\text{min}} \equiv \min \left \{\bankDemand{\price{}; \reserves{t}, \bankBonds{t}, \deposits{t}}, \CARconstraint{\price{}; \reserves{t}, \bankBonds{t}, \deposits{t}, \currencyFraction, \gamma_{t+1}^\text{min}} \right \}.
\label{eq:bankdemand_VaRCAR}
\end{align}
If $\gamma_{t+1}^\text{min} = 0$, then the bank demand~\eqref{eq:bankdemand_VaRCAR} reduces to the original one in Eq.~\eqref{eq:bankdemand_VaR_definition}.

Next we show that capital requirements can act as a stabilizing force in the event of positive shocks, but can also be destabilizing when negative shocks occur. 

\subsection{Stabilizing Effects of Capital Requirements}
\label{sec:stabilizing}

\begin{figure}[htb]
\begin{center}
\includegraphics{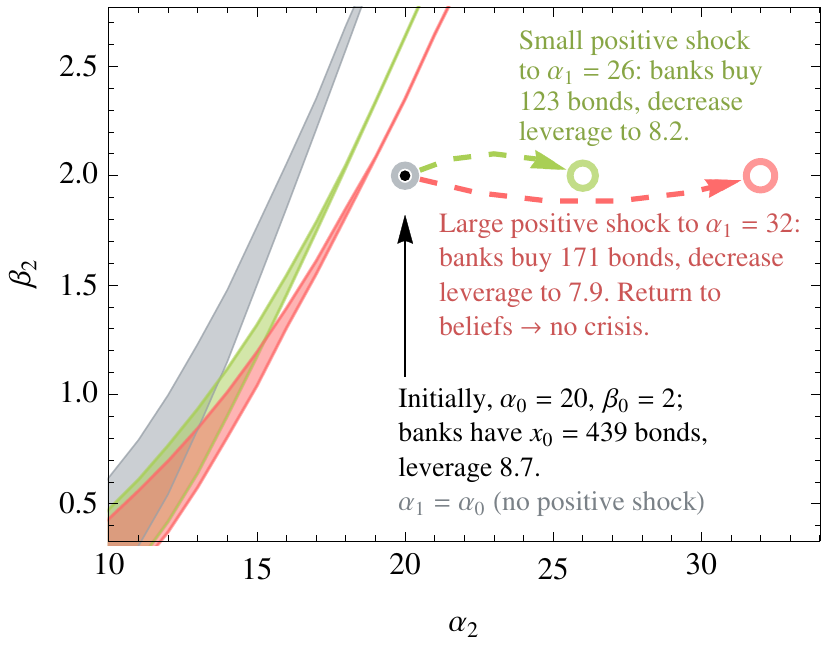}
\caption{Implementing a capital requirement can prevent insolvency when a positive shock is followed by a negative shock. Here, we repeat the experiments in Fig.~\ref{fig:positive_then_negative} with a capital requirement of $\gamma_t^\text{min} \equiv 16\%$ (rather than $\gamma_t^\text{min} \equiv 0$). The banks buy fewer bonds after the positive shocks to $\alpha_1 = 26$ (green circle) and to $\alpha_1 = 32$ (red circle). Consequently, when a second shock occurs, the regions of beliefs  $(\alpha_2, \beta_2)$ that give rise to three equilibria (i.e., the green and red regions) move to the right less than they do when there is no capital requirement, so the system is less vulnerable to insolvency if a negative shock subsequently occurs in period $2$. Unlike in Fig.~\ref{fig:positive_then_negative}, the black dot lies \emph{below} the red region, meaning that restoring the initial beliefs in period $2$ after the large positive shock in period $1$ does not lead to insolvency.}
\label{fig:pos_neg_with_cap_req}
\end{center}
\end{figure}

\begin{figure*}[hbt]
\begin{center}
\includegraphics{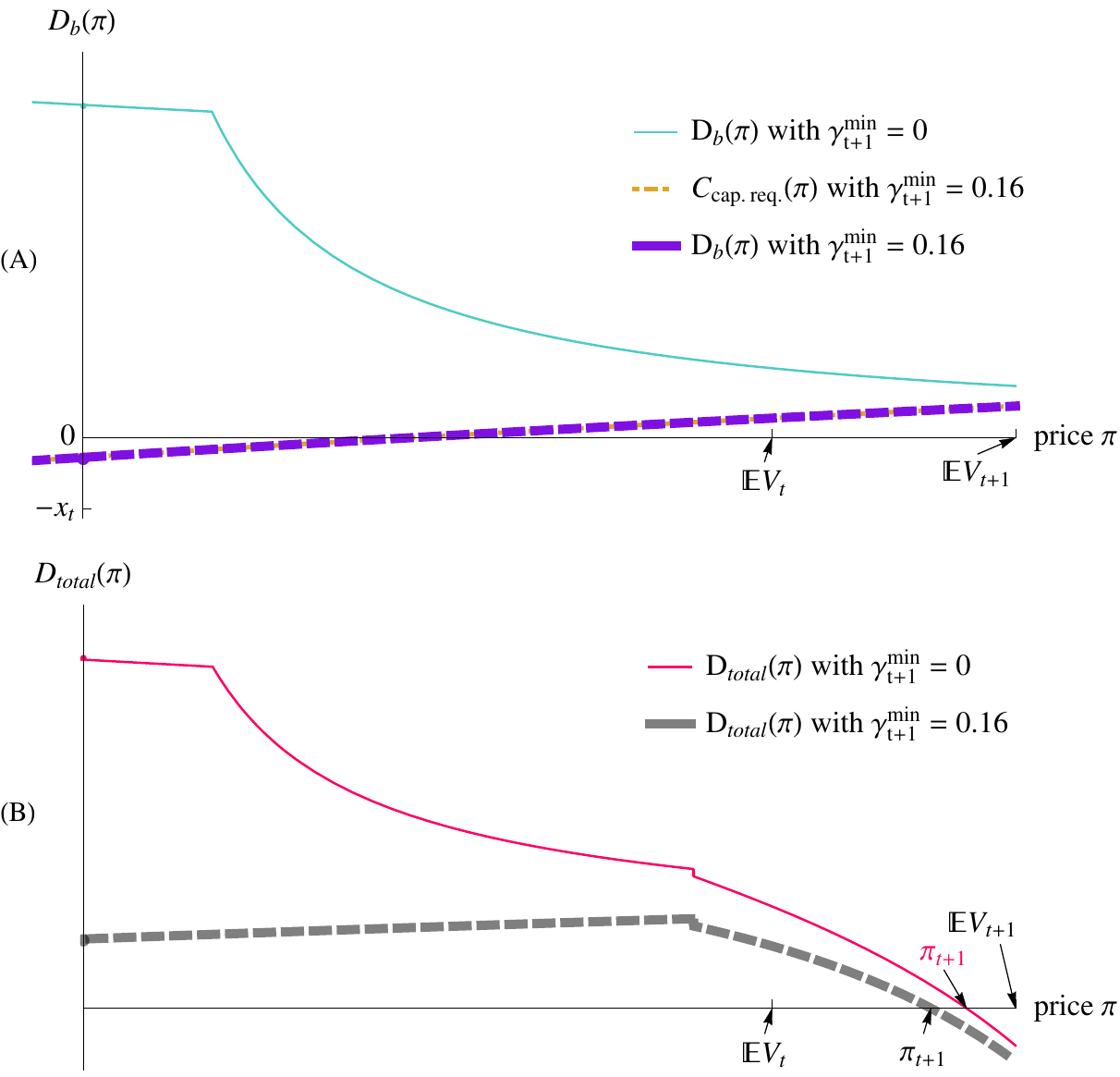}
\caption{When a positive belief shock occurs, a capital requirement can limit the banks' bond purchases. Here, we show the bank demand [panel (A)] and the total demand [panel (B)] for the case of a positive shock to Example~\ref{ex:consistent} from $\alpha_0 = 20$ to $\alpha_1 = 32$ with a capital requirement of $\gamma_t^\text{min} \equiv 16\%$ (i.e., the positive shock in the red case in Fig.~\ref{fig:pos_neg_with_cap_req}). 
The capital requirement binds for all prices shown in panel (A), so the bank demand $\bankDemand{\price{}}$ (purple dashed line) equals the capital constraint $\CARconstraint{\price{}}$ (orange dot-dashed line). Because the banks' demand is reduced in order to satisfy the capital requirement [panel (A)], the total demand is reduced, as shown by the gray dashed line in panel (B). Consequently, the new equilibrium price in the case of a capital requirement of $\gamma_t^\text{min}=16\%$ (labeled $\price{t+1}$ in black) is smaller than the new equilibrium price without a capital requirement (labeled $\price{t+1}$ in pink). Thus, due to the capital requirement, the banks buy fewer bonds, and the bond price increases less, so the system is less vulnerable to insolvency if negative shocks subsequently occur (Fig.~\ref{fig:pos_neg_with_cap_req}).
}
\label{fig:CAR:positive}
\end{center}
\end{figure*}

As noted above, a capital requirement effectively places a limit on banks' bond purchases when a positive shock occurs. An obvious consequence is that it also limits the corresponding growth in banks' leverage and hence their vulnerability to insolvency in the event of a subsequent negative shock. To illustrate this behavior, we repeat the experiments in Fig.~\ref{fig:positive_then_negative} (i.e., a positive shock to Example~\ref{ex:consistent} followed by a negative shock), but this time we implement a capital requirement with $\gamma_t^\text{min} = 16\%$ at every step $t$. (This capital requirement nearly binds in step $0$ because the initial capital-to-assets ratio in Example~\ref{ex:consistent} is $\gamma_0 \approx 16.4\%$, just barely above the minimum of $16\%$.) 
Figure~\ref{fig:pos_neg_with_cap_req} shows the result. 
First, the banks purchase fewer bonds in the positive shocks because of the capital requirement. (Figure~\ref{fig:CAR:positive} shows how the capital requirement affects the demand functions for the severe positive shock in Fig.~\ref{fig:pos_neg_with_cap_req}.)
Because the banks buy fewer bonds in the positive shock, the region of three equilibria does not move as much to the right as they did without a capital requirement (compare the green and red regions in Fig.~\ref{fig:pos_neg_with_cap_req} with those in Fig.~\ref{fig:positive_then_negative}), so the banks are less vulnerable to insolvency in the event of a subsequent negative shock. In fact, the red region now lies \emph{above} the initial belief parameters $(\alpha_0, \beta_0)$, which is marked by a black dot in Fig.~\ref{fig:pos_neg_with_cap_req}. 

Recall that, without a capital requirement, this large positive shock led to an insolvency crisis if it was reverted (Fig.~\ref{fig:positive_then_negative}); by contrast, with a capital requirement of $\gamma_t^\text{min} \equiv 16\%$, reverting this large positive shock does not lead to an insolvency crisis. In other words, our result is supportive of recent arguments for countercyclical capital requirements~\citep{Longworth2010}, which maintain that capital requirements be increased during boom periods and relaxed during crises---a point that is further elaborated in the following section.

\clearpage

\subsection{Destabilizing Effects of Inflexible Capital Requirements}
\label{sec:destabilizing}

Although capital requirements can help banks avoid insolvency in the event of positive-negative shock combinations, inflexible capital requirements can also be problematic.  In particular, for some negative shocks the capital requirement specified in Eqn.~\eqref{eq:CAR:definition} can eliminate the two large-price equilibria (at which the banks would have remained solvent) by forcing the banks to demand to sell even more bonds at those prices. With the two large-price equilibria eliminated, only a small-price equilibrium remains, and the banks are insolvent at this equilibrium. In other words, the capital requirement can force the bond price to collapse and the banks to become insolvent. 

\begin{figure*}[hbt]
\begin{center}
\includegraphics{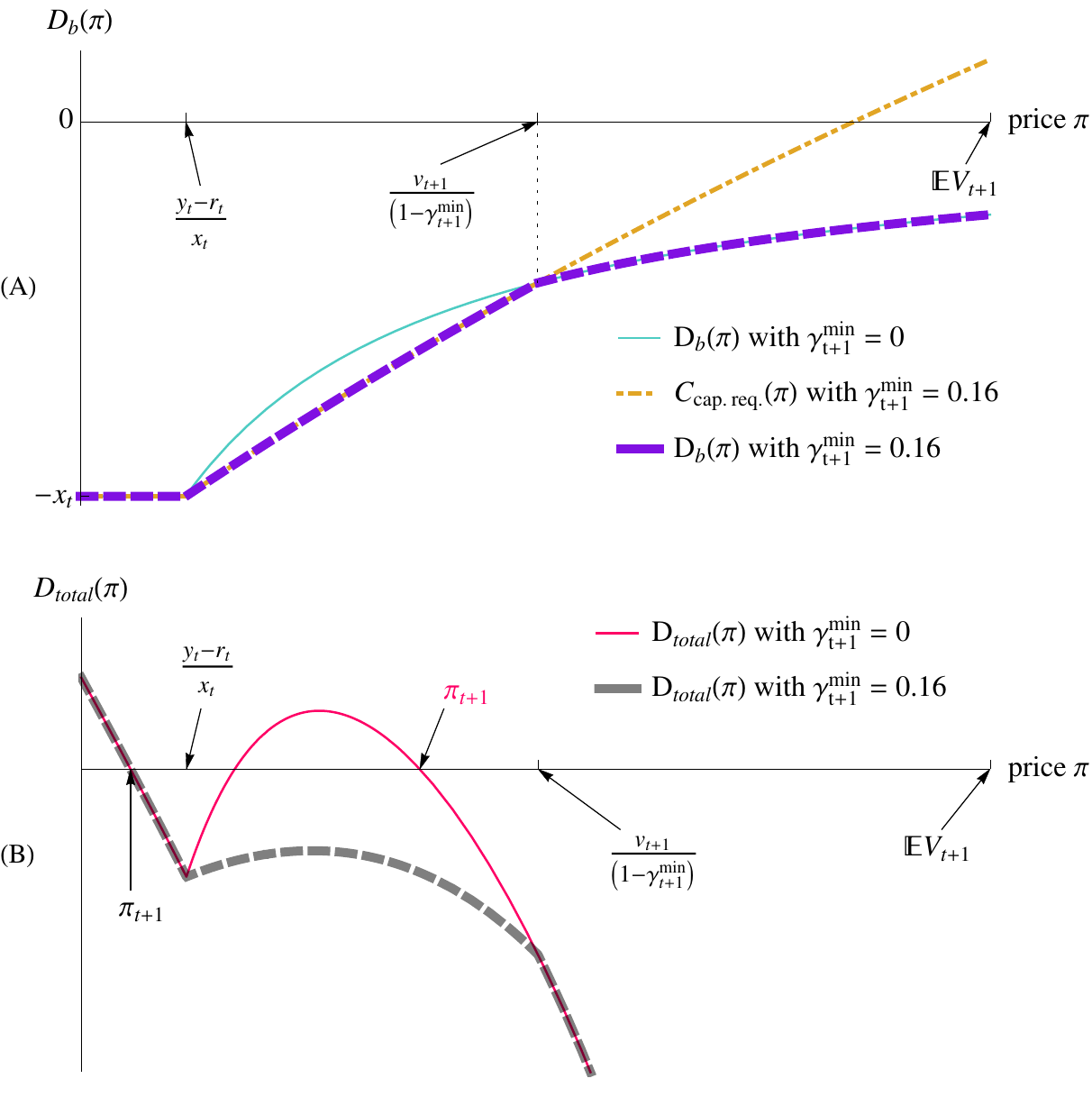}
\caption{When a negative belief shock occurs, a capital requirement can cause the banks to sell bonds even more aggressively, sometimes so much that it causes a large price decline. Here, we illustrate the effect of a capital requirement on the banks' demand function $\bankDemand{\price{}}$ [panel (A)] and on the total demand function $\totalDemand{\price{}}$  [panel~(B)]. The thin lines in panels~(A), (B) are the demand functions for the case of no capital requirement ($\gamma_{t+1}^\text{min} = 0$), while the thick, dashed lines are the demand functions for a capital requirement of $\gamma_{t+1}^\text{min} = 16\%$. Because of the capital requirement~\eqref{eq:CARrule}, the banks' demand cannot exceed $\CARconstraint{\price{}}$ [given by Eq.~\eqref{eq:CAR:definition} and plotted as an orange dot-dashed line in panel (A)]. Consequently, the banks are forced to sell more bonds in order to satisfy the capital requirement if the price $\price{}$ is sufficiently small. Thus, in panel (B) we see that implementing the capital requirement reduces the total demand $\totalDemand{\price{}}$ over a certain range of prices $\price{}$. In some cases, as in the case shown in panel (B), the capital requirement can cause the equilibrium price to decline dramatically: compare the market-clearing price when there is no capital requirement (the horizontal tick labeled $\price{t+1}$ in pink) with the market-clearing price when there is a capital requirement of $16\%$ (the horizontal tick labeled $\price{t+1}$ in black). Further note that this dramatic decline in the bond price causes the banking sector to be insolvent, because $\price{t+1}$ (for $\gamma_{t+1}^\text{min} = 16\%$, marked in black) is smaller than $(\deposits{t}-\reserves{t})/\bankBonds{t}$. The numerical values in this example are a negative shock from $\alpha_0 = 20$ to $\alpha_1 = 16$, with initial condition given by Example~\ref{ex:consistent} (for which $\gamma_0 = 16.4166\%$).
}
\label{fig:CAR:negative}
\end{center}
\end{figure*}

Figure~\ref{fig:CAR:negative} shows an example of such a negative shock that becomes a crisis if there is a capital requirement. Without a capital requirement, there are three equilibrium prices because the total demand function $\totalDemand{\price{}}$ [the pink, thin line in the plot of the bank demand in Fig.~\ref{fig:CAR:negative}(B)] has three roots. Implementing a capital requirement truncates the banks' demand over the prices $[(\deposits{t} - \reserves{t})/\bankBonds{t}, \ve{\epsilon}{t+1}/(1-\gamma_{t+1}^\text{min})]$, as shown in Fig.~\ref{fig:CAR:negative}(A). Consequently, with a capital requirement, the two large-price equilibria have disappeared, leaving only a small equilibrium price that entails bank insolvency [see the unique root of $\totalDemand{\price{}}$ with no capital requirement, given by the thick, dashed curve in Fig.~\ref{fig:CAR:negative}(B)].

\begin{figure}[htb]
\begin{center}
\includegraphics{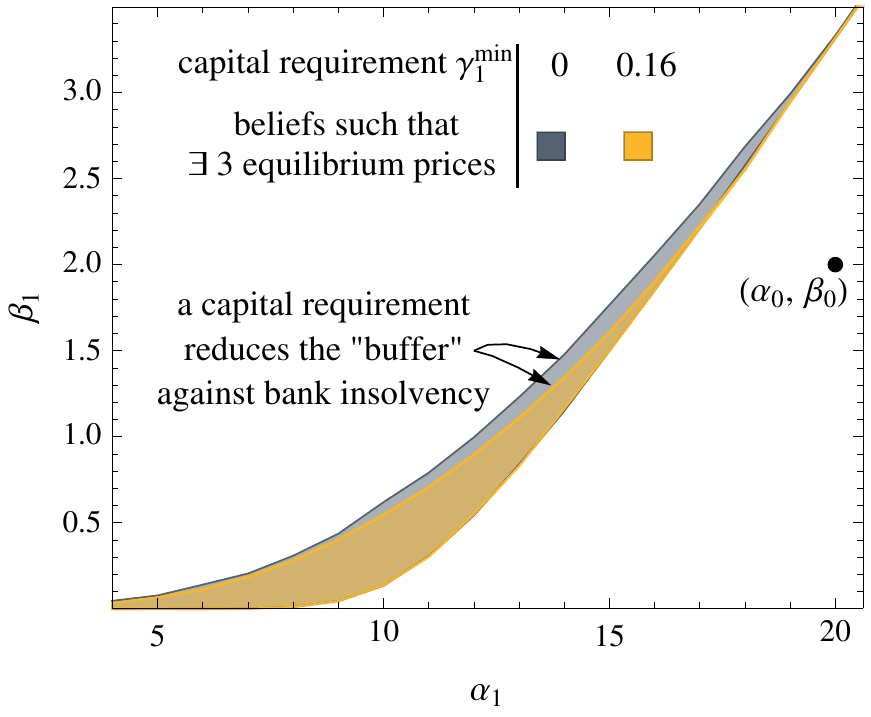}
\caption{Capital requirements can cause bank insolvency in the event of a negative shock. The region of shocked beliefs giving rise to three equilibrium prices (marked by a color region) shrinks from above when one implements a capital requirement. [Here, the initial condition is Example~\ref{ex:consistent}, for which initial capital adequacy ratio $\gamma_0$ is $16.4166\%$, and we implement a capital requirement of $\gamma_1 := 0$ (gray region) or $\gamma_1 := 16\%$ (yellow region).] This region of three equilibria acts as a buffer against bank insolvency, so the capital requirement reduces this buffer. To see this conclusion, consider a pair of belief parameters $(\alpha_1, \beta_1)$ that lie in the gray region but not in the yellow region; if there is a capital requirement, then the bond price declines dramatically in period $1$ and the banks become insolvent, but if there were no capital requirement then no such crisis would occur. 
[Figure~\ref{fig:CAR:negative} shows the demand functions for one such example with $(\alpha_1, \beta_1) = (16,2)$.]}
\label{fig:CAR:negative:regions}
\end{center}
\end{figure}

This event (that a capital requirement causes an insolvency crisis) is rather generic: Fig.~\ref{fig:CAR:negative:regions} shows that implementing a capital requirement shrinks the region of beliefs giving rise to three equilibrium prices. More specifically, a less severe negative shock suffices to move the financial system above the region of three equilibria and into the region of certain bank insolvency. The region of three equilibria acts as a ``buffer'' against bank insolvency, and the capital requirement shrinks that buffer. [For details on why it shrinks the region of three equilibria \emph{from above} (and not from below), as illustrated in Fig.~\ref{fig:CAR:negative:regions},  see~\ref{appendix:cap_req}.]

These destabilizing effects substantiate recent arguments for capital requirements that are graduated or flexible, as argued by~\citet[p.\,189]{Admati2013} and by~\cite{Goodhart2010}. Basel III has taken a step in this direction by implementing two requirements: if a bank's equity lies between $4.5\%$ and $7\%$ of its risk-weighted assets, then the bank is required to slowly rebuild equity by retaining profits and by avoiding paying dividends, but the bank need not raise new equity immediately~\citep{Admati2013}.

\clearpage

\section{Conclusion}
In focusing on inside money and procyclical leverage, our model excludes several other potentially important sources of systemic risk. For example, our assumption that the banks and investors are homogeneous, and hence described by two consolidated balance sheets, effectively rules out knock-on contagion. Asset-to-asset contagion cannot occur because we consider only a single risky asset. Market failure due to information asymmetry cannot occur because of our assumption that the banks and investors share a common belief about the asset's value, and they update their beliefs identically in response to new information. Finally, bank runs cannot occur because we assume that the bank liabilities are implicitly or explicitly guaranteed. 

Although these simplifying assumptions make our model unrealistic in many respects, they also highlight how little complexity is needed to generate systemic risk; specifically, the combination of inside money and procyclical leverage suffices to generate crises. Moreover, the model provides a framework for understanding the effects of policies such as leverage constraints and emergency asset price supports. Many of the historically unprecedented actions taken by the Federal Reserve in 2008 involved attempts to support asset prices that the central bank could not directly purchase. By creating facilities that accepted these assets as collateral in exchange for loans of reserves or Treasury securities, the Fed may have been trying to coordinate beliefs on more optimistic equilibria with a solvent banking sector. This action was a response to crisis conditions that arose in part because of excessive leverage, which countercyclical capital requirements could have held in check. Our model demonstrates the need for both types of policy in order to mitigate or prevent the catastrophic consequences of a reversal of optimism.

\newpage

\appendix

\section{Properties of the investor demand function}
\label{sec:investordemand}

Here, we formalize and prove properties of the investors' demand function that were mentioned in the text. Lemma~\ref{lem:Dn} shows that the investor demand is single-valued, under the following mild assumptions on the investors' utility function and on the random variable representing beliefs about the bonds' value. Next, Lemma~\ref{lem:CRRA} achieves stronger results for the particular utility function with constant relative risk aversion (CRRA).

Lemma~\ref{lem:Dn}--\ref{lem:CRRA} make the following assumptions. (All four assumptions except for Assumption~\ref{as:C2} and the second part of Assumption~\ref{as:cts_density} were mentioned in the main text.)

\begin{assumption}
The investors' utility function $u : (0, \infty) \to \mathbb{R}$ is strictly increasing and strictly concave.
\label{as:increasing_concave}
\end{assumption}

\begin{assumption}
The investors' utility function $u(\cdot)$ is twice continuously differentiable. 
\label{as:C2}
\end{assumption}

\begin{assumption}
$V_{t+1}$ is an absolutely continuous random variable, and its density function $f_{t+1}$ is a continuous function on $[0,1]$. 
\label{as:cts_density}
\end{assumption}

\begin{assumption}
The investors have some bonds $(\investorBonds{t} > 0)$, some deposits $(\deposits{t}>0)$, and some currency $(\currency{t}>0)$.
\label{as:positive}
\end{assumption}

Some elementary results in convex optimization imply that the investors' demand is single-valued and that they want to sell all their bonds if and only if the price equals or exceeds the expected value of bonds. 

\begin{lemma}[Investor demand is single-valued and equals $-\investorBonds{t}$ iff price $\price{} \geq \E V_{t+1}$]
Under Assumptions~\ref{as:increasing_concave}--\ref{as:positive}, the investor demand function is single-valued. 
Furthermore, $\investorDemand{\price{} }= -\investorBonds{t}$ if and only if $\price{} \geq \E V_{t+1}$. 
\label{lem:Dn}
\end{lemma}
\begin{proof}
Let $\investorBonds{t}, \deposits{t}, \currency{t}>0$ (by Assumption~\ref{as:positive}), and let the currency fraction $\currencyFraction \in [0,1)$ 
and the hypothetical price $\price{} \in (0,1)$. 
Recall that the investors' demand, Eq.~\eqref{eq:nonbankdemand} of the main text, is defined to be 
\begin{align}
& \investorDemand{\price{}; \currency{t}, \deposits{t}, \investorBonds{t}} :=  \argmax_{-\investorBonds{t} \, \leq \, d \, \leq \, \deposits{t}/[\price{} (1-\currencyFraction)]} \E \left ( u \left [ V_{t+1}(\investorBonds{t}+d) + \deposits{t} + \currency{t} - \price{} d \right ] \right ). \label{eq:investordemand_copied} 
\end{align}
For convenience, let $g(d; \investorBonds{t}, \deposits{t}, \currency{t}, \price{})$ denote the objective function of $\investorDemand{\price{}}$, i.e.,
\begin{align*}
g(d; \investorBonds{t}, \deposits{t},  \currency{t}, \price{}) &:=  \E \left ( u \left [ V_{t+1}(\investorBonds{t}+d) + \deposits{t} + \currency{t} - \price{} d \right ] \right ) \\
&=   \int_0^1 u \left [ v(\investorBonds{t} + d) + \deposits{t} + \currency{t} - \price{} d \right ] f_{t+1}(v) dv,
\end{align*}
The domain of $g$ is the interval $[-\investorBonds{t}, \deposits{t}/(\price{} (1-\currencyFraction)) ]$, which is convex.

The continuity of $u$ and $u'$ (by Assumption~\ref{as:C2}) and of $f_{t+1}$ (by Assumption~\ref{as:cts_density}) enable us to use the Leibniz integration rule twice to move the derivative inside the integral sign to compute the first two derivatives 
\begin{subequations}
\begin{align}
\frac{\partial g}{\partial d}
&= \int_0^1 u' \left [ v(\investorBonds{t} + d) + \deposits{t} + \currency{t} - \price{} d \right ] (v-\price{}) f_{t+1}(v) dv,
\label{eq:dobjfn}
\end{align}
\begin{align}
\frac{\partial^2 g}{\partial d^2} 
&= \int_0^1 u'' \left [ v(\investorBonds{t} + d) + \deposits{t} + \currency{t} - \price{} d \right ] (v-\price{})^2 f_{t+1}(v) dv.
\label{eq:d2objfn}
\end{align}
\end{subequations}

Because $u$ is strictly concave (by Assumption~\ref{as:increasing_concave}) and twice differentiable (by Assumption~\ref{as:C2}), and because its domain $(0,\infty)$ is convex, we know that $u'' < 0$~\cite[Sec.\ 3.1.4, page 71]{Convex_Optimization}. 
Also, we know that $(v-\price{})^2 \geq 0$ with equality if and only if $v = \price{}$. Combining these two conclusions with Eq.~\eqref{eq:d2objfn} gives
\begin{align}
\frac{\partial^2 g}{\partial d^2}  < 0 \quad \text{ for all } d \in 
\left [-\investorBonds{t}, \frac{\deposits{t}}{\price{} (1-\currencyFraction)} \right ].\label{eq:negative_second_derivative}
\end{align}
Because $\partial^2 g/\partial d^2 < 0$ and because the domain of $g$ is convex, we know that $g(d)$ is strictly concave~\cite[Sec.\ 3.1.3, page 69]{Convex_Optimization}. Thus, any solution $d^*$ to the first-order (necessary) condition for the maximization in Eq.~\eqref{eq:investordemand_copied},
\begin{align}
\frac{\partial g}{\partial d}(d^*; \investorBonds{t}, \deposits{t},  \currency{t}, \price{}) = 0,
\label{eq:FO}
\end{align}
is a unique, global maximum~\cite[Sec.\ 3.1.3, page 69]{Convex_Optimization}. If no solution $d^*$ to Eq.~\eqref{eq:FO} exists, then $\partial g / \partial d$ is either positive for all $d$, in which case $g(d)$ has a unique maximum at $-\investorBonds{t}$, or $\partial g / \partial d$ is negative for all $d$, in which case $g(d)$ has a unique maximum at $\deposits{t} / [\price{} (1-\currencyFraction)]$. Thus, the investor demand function~\eqref{eq:investordemand_copied} is single-valued.

Furthermore, the first derivative of the objective function evaluated at the lower constraint $d=-\investorBonds{t}$ is
\begin{align*}
\frac{\partial g}{\partial d}  (-\investorBonds{t}; \investorBonds{t}, \deposits{t}, \currency{t}, \price{}) &= (\E V_{t+1} - \price{}) u'(\deposits{t} + \currency{t} + \price{} \investorBonds{t}).
\end{align*}
Because $u'>0$ (by Assumptions~\ref{as:increasing_concave}--\ref{as:C2}), we know that 
\begin{align}
\frac{\partial g}{\partial d}  (-\investorBonds{t}; \investorBonds{t}, \deposits{t}, \currency{t}, \price{}) \leq 0 \quad \text{ if and only if } \quad \price{} \geq \E V_{t+1}
\label{eq:slope_negative}
\end{align}
Equations~\eqref{eq:slope_negative} and~\eqref{eq:negative_second_derivative} imply that $\investorDemand{\price{}} = -\investorBonds{t}$ if and only if $\price{} \geq \E V_{t+1}$, which completes the proof. 
\end{proof}

The next lemma achieves stronger results for the particular utility function $u(\cdot)$ that exhibits constant relative risk aversion. Specifically, the lemma establishes the price at which the investors first begin to buy less than the maximum that they can afford.

\begin{lemma}[Investor demand less than they can afford iff $\price{} > \priceDemandLessThanMax_{t+1}$]
Suppose that the investors have CRRA utility with parameter $\lambda$ and that the belief follows a Beta distribution with parameters $\alpha_t, \beta_t$. Then under Assumptions~\ref{as:increasing_concave}--\ref{as:positive}, we know that $\price{} \leq \priceDemandLessThanMax_{t+1}$ implies $\investorDemand{\price{}} = \deposits{t}/[\price{}(1-\currencyFraction)]$, and $\priceDemandLessThanMax_{t+1} < \price{} \leq \E V_{t+1}$ implies $-\investorBonds{t} \leq \investorDemand{\price{}} < \deposits{t}/[\price{}(1-\currencyFraction)]$.
\label{lem:CRRA}
\end{lemma}

\begin{proof}
We will show that the first derivative with respect to $d$ of the objective function $g(d; \investorBonds{t}, \deposits{t}, \currency{t}, \price{})$ of the maximization in $\investorDemand{\price{}}$ evaluated at the upper constraint $d = \deposits{t}/[\price{} (1 - \currencyFraction)]$ is negative for $\price{} < \priceDemandLessThanMax_{t+1}$ and positive for $\price{} > \priceDemandLessThanMax_{t+1}$. By Lemma~\ref{lem:Dn}, we know that $\partial^2 g(d; \investorBonds{t}, \deposits{t}, \currency{t}, \price{}) / \partial d^2 < 0$ for all $-\investorBonds{t} \leq d \leq \deposits{t}/[\price{} (1 - \currencyFraction)]$. Because of this negative second derivative, we know that 
\begin{itemize} 
\item $\partial g / \partial d \leq 0$ at $d = \deposits{t}/[\price{} (1 - \currencyFraction)]$ implies $\investorDemand{\price{}} = \deposits{t}/[\price{} (1 - \currencyFraction)]$, and that 
\item $\partial g / \partial d > 0$ at $d = \deposits{t}/[\price{} (1 - \currencyFraction)]$ implies $\investorDemand{\price{}} < \deposits{t}/[\price{} (1 - \currencyFraction)]$,
\end{itemize}
which proves the claim. 

Using $u'(w) = w^{-\lambda}$ in Eq.~\eqref{eq:dobjfn} and evaluating at $d=\deposits{t}/[\price{} (1 - \currencyFraction)]$ gives
\begin{align*}
\left. 
\frac{\partial}{\partial d} g(d; \investorBonds{t}, \deposits{t}, \currency{t}, \price{}) \right |_{d=\deposits{t}/[\price{} (1 - \currencyFraction)]} 
&= \int_0^1 u' \left [ v \left (\investorBonds{t}+ \frac{\deposits{t}}{\price{} (1-\currencyFraction)} \right ) \right ] (v-\price{}) f_{t+1}(v) dv \\
\quad &= \left [\investorBonds{t} + \frac{\deposits{t}}{\price{}(1-\currencyFraction)} \right ]^{-\lambda} \int_0^1 \left ( v^{1-\lambda} - \price{} v^{-\lambda} \right ) f_{t+1}(v) dv \\
\quad &= \left [\investorBonds{t} + \frac{\deposits{t}}{\price{}(1-\currencyFraction)} \right ]^{-\lambda} \left ( \E \left [ (V_{t+1})^{1-\lambda} \right ] - \price{} \E \left [ (V_{t+1})^{-\lambda} \right ] \right ),
\end{align*}
which is positive if and only if 
\begin{align*}
\price{} > \priceDemandLessThanMax_{t+1} \equiv \frac{\E \left [ (V_{t+1})^{1-\lambda} \right ]}{\E \left [ (V_{t+1})^{-\lambda}\right ]}  = \frac{\alpha_{t+1} - \lambda}{\alpha_{t+1} + \beta_{t+1} - \lambda}
\end{align*}
because $\investorBonds{t} + \deposits{t} / [\price{}(1-\currencyFraction)] > 0$ by Assumption~\ref{as:positive}. This equivalence proves the claim.
\end{proof}

\section{Derivation of the bank demand function}
\label{sec:derive_bank_demand}

Here we derive the bank demand function for positive shocks [Eq.~\eqref{eq:bankDemand:positiveShock}] and for negative shocks [Eq.~\eqref{eq:bankDemand:negativeShock}] from the original definition [Eq.~\eqref{eq:bankdemand_VaR_definition}]. 
By rearranging the insolvency constraint, Eq.~\eqref{eq:solvencyequivalence}, we find that the three constraints~\eqref{eq:bankDemandConstraints} are equivalent to 
\begin{subequations}
\begin{align}
-\bankBonds{t} \leq d \leq \min \left \{ \frac{ \ve{\epsilon}{t+1} \bankBonds{t} + \reserves{t} - \deposits{t}}{\price{} -  \ve{\epsilon}{t+1}}, \frac{\reserves{t}}{\currencyFraction \price{}} \right \} \qquad & \text{if } \price{} \geq \ve{\epsilon}{t+1};
\label{eq:constraintequivalence} \\
\max \left \{ -\bankBonds{t}, \frac{ \ve{\epsilon}{t+1} \bankBonds{t} + \reserves{t} - \deposits{t}}{ \price{} - \ve{\epsilon}{t+1}} \right \} \leq d \leq \frac{\reserves{t}}{\currencyFraction \price{}} \qquad & \text{if } \price{} < \ve{\epsilon}{t+1}.
\label{eq:constraintequivalenceSmallerThanVaR}
\end{align}
\label{eq:bankDemandConstraintsRewritten}
\end{subequations}

To evaluate the 
$\argmax$ 
in Eq.~\eqref{eq:bankdemand_VaR_definition}, note that the marginal change in the banks' expected equity due to a infinitesimal increase in demand $d$ is
\begin{align*}
\frac{\partial}{\partial_d} \E \bankEquity{t+1} = 
\frac{\partial}{\partial_d} \E \left [ V_{t+1} (\bankBonds{t}+d) + \reserves{t} - (\deposits{t}+\price{} d) \right ] = \E V_{t+1} - \price{}.
\end{align*}
Thus, the expected equity $\E \bankEquity{t+1}$ is linear in $d$ with slope $\E V_{t+1} - \price{}$, subject to the constraint~\eqref{eq:bankDemandConstraintsRewritten}. 
The sign of $\E V_{t+1} - \price{}$ therefore determines whether the bank demand is the lower or upper constraint in~\eqref{eq:bankDemandConstraintsRewritten}, and the sign of $\pi - \ve{\epsilon}{t+1}$ determines whether to use constraint~\eqref{eq:constraintequivalence} or~\eqref{eq:constraintequivalenceSmallerThanVaR}. Note that if $\E V_{t+1} < \price{} < \ve{\epsilon}{t+1}$, then the banks' demand is the lower constraint in~\eqref{eq:constraintequivalenceSmallerThanVaR}, which simplifies to $-\bankBonds{t}$ under the assumption that $\price{} \bankBonds{t} + \reserves{t} - \deposits{t} \geq 0$. 
In summary, the banks' demand function can be written more explicitly as $\bankDemand{\price{}} = -\bankBonds{t}$ if $\price{} \bankBonds{t} + \reserves{t} - \deposits{t} < 0 \text{ or if } \price{} > \E V_{t+1}$, and otherwise
\begin{align}
\bankDemand{\price{}} &= 
\begin{dcases}
     \frac{\reserves{t}}{\currencyFraction  \price{}} & \text{if } 
     \price{} < \ve{\epsilon}{t+1} \\
 \min \left \{ \frac{ \ve{\epsilon}{t+1} \bankBonds{t} + \reserves{t} - \deposits{t}}{\price{} -  \ve{\epsilon}{t+1}}, \frac{\reserves{t}}{\currencyFraction \price{}} \right \}  & \text{else} 
\end{dcases}.
\label{eq:bankdemand_VaR_rewritten} 
\end{align}
Finally, considering whether the banks comply with their insolvency constraint~\eqref{eq:solvencyequivalence} immediately after the shock leads to Eq.~\eqref{eq:bankDemand_shocks}.

\section{Effect of a capital requirement on the region of beliefs giving rise to three equilibria}
\label{appendix:cap_req}
Here we explain why, after implementing a capital requirement, the region of beliefs giving rise to three equilibria is reduced from above (and not from below), as illustrated in Fig.~\ref{fig:CAR:negative:regions}. 
Recall that the banks are insolvent if and only if the bond price $\price{} \leq (\deposits{t}-\reserves{t})/\bankBonds{t}$. 
Also recall [from the banks' demand in the event of a negative shock, illustrated in
Fig.~\ref{fig:bankdemand}(B)] that below the price $(\deposits{t}-\reserves{t})/\bankBonds{t}$ the banks are forced to sell all their bonds; by contrast, above the price $(\deposits{t}-\reserves{t})/\bankBonds{t}$ and below $\E V_{t+1}$, the banks' demand increases with the price. That is, the kink in the bank demand occurs at the price $(\deposits{t}-\reserves{t})/\bankBonds{t}$. 

Implementing a capital requirement does not affect the location of this kink because the capital constraint $\CARconstraint{\price{}}$ [defined in Eq.~\eqref{eq:CAR:definition}] satisfies $\CARconstraintBracket{(\deposits{t}-\reserves{t})/\bankBonds{t}} = -\bankBonds{t}$. 
If deposits $\deposits{t}$ exceed reserves $\reserves{t}$ (as typically occurs in practice), then $\CARconstraint{\price{}}$ is increasing in the price $\price{}$ for $\price{} \geq (\deposits{t}-\reserves{t})/\bankBonds{t}$, so a kink still occurs at the price $(\deposits{t}-\reserves{t})/\bankBonds{t}$. 
On the other hand, if reserves $\reserves{t}$ exceed deposits $\deposits{t}$ (which rarely occurs in practice), then $\CARconstraint{\price{}} > 0$ for all prices $\price{}$, so the capital requirement does not bind for any price because the bank's demand is negative for a negative shock, and so the bank demand still has a kink at the price $(\deposits{t}-\reserves{t})/\bankBonds{t}$. 
In summary, the position of the first saddle-node bifurcation (at which two new equilibrium prices appear) is $(\deposits{t}-\reserves{t})/\bankBonds{t}$ for any minimum capital-to-assets ratio $\gamma_{t+1}^\text{min}$; that is, the lower boundary of the region of three equilibria [such as in Fig.~\ref{fig:negativeshock}(E) and in Fig.~\ref{fig:CAR:negative:regions}] is independent of the capital requirement. 

Although a capital requirement does not move the location of the kink in the bank demand, a capital requirement can bind (and hence reduce the banks' demand) for prices just above that price where the kink occurs, $(\deposits{t}-\reserves{t})/\bankBonds{t}$, as illustrated in Fig.~\ref{fig:CAR:negative}(A). Consequently, the left-hand side of the ``hump'' in the total demand function [depicted in Fig.~\ref{fig:negativeshock}(A)--(C)] is truncated; for an illustration, see Fig.~\ref{fig:CAR:negative}(A). Thus, a less severe negative shock causes the two larger equilibrium prices to disappear, as illustrated in Fig.~\ref{fig:CAR:negative}(B). In summary, the reduction in the banks' demand just above the price $(\deposits{t}-\reserves{t})/\bankBonds{t}$ explains why the region of three equilibria is truncated from above in Fig.~\ref{fig:CAR:negative:regions} and hence why the capital requirement can force a decline in the bond price and bank insolvency.

\newpage 


\bibliographystyle{elsarticle-harv}


\end{document}